\documentstyle[preprint,aps,floats,epsfig]{revtex}
\tightenlines

\newcommand{\ie}{i.e.}
\newcommand{\eg}{e.g.}
\def\appendix{\par
 \setcounter{section}{0}
 \setcounter{subsection}{0}
 \def\thesection{Appendix \Alph{section}}
 \def\thesubsection{\Alph{section}.\arabic{subsection}}
 \def\theequation{\Alph{section}.\arabic{equation}}
 \setcounter{equation}{0}}

\begin{document}

\renewcommand{\thefootnote}{\fnsymbol{footnote}}

\title{Higgs-mediated flavor-changing neutral currents\\
 in the general framework with two Higgs doublets\\
 -- an RGE analysis}

\begin{flushright}
hep-ph/9806282\\
June 5, 1998\\
updated version Oct. 28, 1998 --\\
Jour. ref.: Phys. Rev. D {\bf 58}, 116003 (1998)
\end{flushright}

\centerline{{\large \bf
Higgs-mediated flavor-changing neutral currents}}
\centerline{{\large \bf 
in the general framework with two Higgs doublets}}
\centerline{{\large \bf
 -- an RGE analysis}}

\vspace{0.8cm}

\centerline{ G.~Cveti\v c\footnote[1]{e-mail:
cvetic@doom.physik.uni-dortmund.de}}

\centerline{{\it  
Department of Physics, Universit\"at Dortmund,
44221 Dortmund, Germany}}

\vspace{0.5cm}

\centerline{ S.S.~Hwang\footnote[2]{e-mail:
sshwang@theory.yonsei.ac.kr} and C.S.~Kim\footnote[3]{e-mail:
kim@cskim.yonsei.ac.kr; http://phya.yonsei.ac.kr/\~{}cskim/}}

\centerline{{\it Department of Physics, Yonsei University, 
Seoul 120-749, Korea}} 

\renewcommand{\thefootnote}{\arabic{footnote}}

\begin{abstract}

We consider the standard model with two Higgs doublets with 
the most general Yukawa coupling terms (``type III''). 
In the model, the neutral-Higgs-mediated flavor-changing neutral 
currents (FCNC's) are allowed, but must be reasonably suppressed
at low energies of probes. It has been known that the 
existing hierarchies of quark masses render this suppression
at low energies rather natural. On the other hand, the
model has been regarded by many as unnatural because of
the absence of any symmetry that would ensure persistence
of this suppression as the energy of probes increases.
The opinion has been based on the expectation that the mentioned 
FCNC's would increase by large factors at increasing energies. 
We performed a numerical analysis of the flow of these FCN 
coupling parameters as governed by the one-loop renormalization 
group equations (RGE's), in a simplified case when Yukawa 
couplings of the first quark generation are neglected. The 
analysis shows a remarkable persistence of the mentioned FCNC 
suppression and thus indicates that the model is not unnatural 
in the RGE sense. Further, we point out two mistakes in the 
Yukawa RGE's of Machacek and Vaughn at one-loop level.\\
PACS number(s): 11.10.Hi; 12.15.Mm; 12.38.Bx; 12.60.Fr

\end{abstract}

\setcounter{equation}{0}
\newpage

\section{Introduction}

The (standard) model with two Higgs doublets whose Yukawa
couplings in the quark sector have the most general form
was apparently first introduced already in 1973 by
T.~D.~Lee \cite{Lee}. His main motivation for introducing
the model, later also known as the general two-Higgs-doublet
model (G2HDM) or ``type III'' 2HDM, 
was to study new possible CP-violating phenomena.
Others \cite{Sikivieetal} continued investigating 
phenomenology of the model along these lines.

Subsequently, Glashow and Weinberg \cite{GlashowWeinberg}
in 1977 stressed that only those models with two 
Higgs doublets whose Yukawa coupling sector possesses
specific discrete [or equivalently U(1)-type]
family symmetries lead to automatic and full suppression
of the effective flavor-changing neutral (FCN) Yukawa 
parameters, and ensure this suppression at any energy of 
probes. They pointed out that there are basically two
types of such 2HDM's -- so called ``type I'' and
``type II'' models, in which either one Higgs doublet 
alone is responsible for all the
quark masses [2HDM(I)], or one Higgs doublet is
responsible for all the up-type quark masses and the other
for the down-type quark masses [2HDM(II)]. This point
of Glashow and Weinberg apparently had a great impact on
the physics community, especially because most of the
mentioned flavor-changing neutral currents (FCNC's)
mediated by neutral Higgs\footnote{
A more precise expression would be ``neutral flavor-changing
scalar (Yukawa) coupling,'' since these couplings
have no four-vector current structure involving $\gamma^{\mu}$.} 
must be strongly suppressed
at low energies of probes ($\stackrel{<}{\sim}\!E_{\rm ew}$)
due to the firm experimental evidence of FCNC suppression.
Consequently, the general 2HDM, which has no such automatic 
and full suppression of the FCN Yukawa coupling parameters, 
apparently wasn't investigated by physicists until the late 
eighties.

Since the late eighties, there has been a moderate resurgence
of investigation of the G2HDM \cite{Brancoetal}-\cite{Xuetal}.
These works investigate {\em low\/} energy phenomena
($E\!\stackrel{<}{\sim}\!E_{\rm ew}$) as predicted by
G2HDM's, with most of the FCN Yukawa coupling parameters 
(at low energies of probes) being generally nonzero 
but reasonably suppressed.\footnote{
Low energy experiments show that those flavor-changing neutral 
coupling parameters which don't involve $t$ quark
are suppressed in nature at low energies $E\!\sim\!m_q$,
while for those involving $t$ quark there is no experimental 
evidence yet available.}
Refs.~\cite{Brancoetal} investigate predictions of the
model mainly for CP-violating,
and Refs.~\cite{Xuetal} mainly for FCNC-violating
phenomena. The resulting amplitudes then include
FCN Yukawa coupling parameters at low energies of probes. 

The authors Cheng, Sher and Yuan (CSY) \cite{CSY}
offered arguments which render the G2HDM reasonably
natural from the aspect of {\em low\/} energy physics,
thus countering one part of the reservations based on
the arguments of Glashow and Weinberg \cite{GlashowWeinberg}.
CSY basically proposed specific ans\"atze for the 
Yukawa parameters in the
G2HDM at low energies of probes, specifically the FCN
Yukawa coupling parameters,
motivated laregly by the existing mass hierarchy
of quarks. Therefore, their ans\"atze are reasonably
natural, or, more conservatively, not ``unnatural.''
Motivation of their ans\"atze didn't explicitly
involve any family symmetries.
Moreover, they showed that their ans\"atze allow
the masses of neutral scalars to be as low as
$\sim\!10^2$ GeV while still not violating the
available (low energy) data on suppressed FCNC phenomena.
Later on, Antaramian, Hall and Ra\v{s}in (AHR) \cite{AHR} 
proposed somewhat similar (but not identical)
ans\"atze, which they motivated by their requirement
that the Yukawa interactions have certain approximate
flavor symmetries.
The CSY and similar ans\"atze were mainly used 
by other authors \cite{Brancoetal} and \cite{Xuetal}
in their investigations of low energy phenomenology
of the G2HDM.

We wish to reemphasize that the mentioned ans\"atze
countered only {\em one part\/} of the arguments
(based on \cite{GlashowWeinberg}) against the G2HDM.
The symmetry arguments of Glashow and Weinberg 
\cite{GlashowWeinberg} didn't just suggest that a
natural 2HDM should have a well motivated suppression
of the flavor-changing neutral (FCN) Yukawa parameters
at {\em low\/} energies, but that these FCN parameters
should remain suppressed also when the energy of probes
increases. Based on this latter point of \cite{GlashowWeinberg},
a large part of physics community has continued regarding the
G2HDM as unnatural. The main point against the G2HDM has
consisted of the fear, or conjecture, that the FCN
Yukawa coupling parameters in the G2HDM,
even though suppressed at low energies by reasonably motivated
arguments, would behave unnaturally as the energy of probes
increases. Stated otherwise, it has been expected that at least
some FCN Yukawa coupling parameters 
would increase by a large factor or even by
orders of magnitude at increased energies (well below the
Landau pole), due to the absence of explicit discrete
[or U(1)-type] family symmetries in the Yukawa sector.
The absence of such symmetries, so the argument,
would in general result in a strong ``pull-up'' effect on
the small flavor-changing by the much larger flavor-conserving
Yukawa coupling parameters as the evolving energy increases.
In such a case, the model would then generally 
contain a (thus unnatural) fine-tuning: 
large ``bare'' FCN Yukawa coupling parameters at high energies 
would have to be fine-tuned
in order to obtain at low energies their phenomenologically
acceptable suppression.

Therefore, we investigate this question in the present paper,
by performing a numerical analysis of the one-loop
renormalization group equations (RGE's) of the G2HDM.
In Sec.~2 we present the model and write down conditions
for the suppression of FCN Yukawa coupling parameters 
at low energies (CSY ansatz).
In Sec.~3 we write down the one-loop RGE's for the
Yukawa coupling parameters in the G2HDM in a specific
form convenient for numerical analyses. A short derivation
of the RGE's is given in the Appendix. Sec.~3 contains also
comparison with the existing literature on RGE's.
In Sec.~4 we then numerically investigate the
RGE evolution of the Yukawa coupling parameters for quarks, 
neglecting for simplicity the Yukawa parameters of the light
first generation of quarks. We present the resulting
evolutions of the FCN Yukawa coupling parameters 
for various low energy ans\"atze, i.e., essentially for 
variations of the CSY ans\"atze. We also observe some
other interesting properties of the presented evolution.
Sec.~5 contains a summary and conclusions.

\section{The model and low energy ans\"atze}

Yukawa interactions for quarks in the G2HDM in any 
$\rm{SU(2)_L}$-basis have the most general form
\begin{eqnarray}
\lefteqn{
{\cal L}^{(E)}_{\rm G2HDM} =
- \sum_{i,j=1}^3 {\Big \lbrace}
{\tilde D}_{ij}^{(1)}( {\bar {\tilde q}}^{(i)}_L {\Phi}^{(1)} )
{\tilde d}^{(j)}_{R} +
{\tilde D}_{ij}^{(2)}( {\bar {\tilde q}}^{(i)}_L {\Phi}^{(2)} )
{\tilde d}^{(j)}_{R} + }
    \nonumber\\
& & + {\tilde U}_{ij}^{(1)}( {\bar {\tilde q}}^{(i)}_L \tilde {\Phi}^{(1)} )
{\tilde u}^{(j)}_{R} +
{\tilde U}_{ij}^{(2)}( {\bar {\tilde q}}^{(i)}_L \tilde {\Phi}^{(2)} )
{\tilde u}^{(j)}_{R} + {\mathrm{h.c.}}
{\Big \rbrace}
+ \lbrace  {\bar \ell} {\Phi} {\ell}{\mathrm{-terms}}  \rbrace \ .
\label{2HD30}
\end{eqnarray}
The tildes above the Yukawa coupling parameters
and above the quark fields mean that these quantities are 
in an arbitrary $\rm{SU(2)_L}$-basis (i.e., weak basis,
not the mass basis).
The superscript $(E)$ at the Lagrangian density means that
the theory has a finite effective energy cutoff $E$,
and the reference to this evolution energy
$E$ was omitted at the fields and at the
Yukawa coupling parameters in order to have simpler notation
($E\!\sim\!10^2$ GeV for renormalized quantities). 
The following notations are used:
\begin{equation}
{\Phi}^{(k)}  \equiv  { {\phi}^{(k)+} \choose {\phi}^{(k)0} } 
\equiv \frac{1}{\sqrt{2}} 
{ {\phi}_1^{(k)} + {\mathrm{i}} {\phi}_2^{(k)} \choose
     {\phi}_3^{(k)} + {\mathrm{i}} {\phi}_4^{(k)} }  \ , 
\label{2HDnot0}
\end{equation} 
\begin{equation}
{\tilde {\Phi}}^{(k)}  \equiv {\mathrm{i}} {\tau}_2 
{\Phi}^{ (k) \dagger {\rm T} } 
\equiv \frac{1}{\sqrt{2}} 
{ {\phi}_3^{(k)} - {\mathrm{i}} {\phi}_4^{(k)} \choose
     -{\phi}_1^{(k)} + {\mathrm{i}} {\phi}_2^{(k)} }  \ , 
\label{2HDnot1}
\end{equation}
\begin{equation}
{\tilde q^{(i)}} = { {\tilde u^{(i)}} \choose {\tilde d^{(i)}} } \ :
\qquad
{\tilde q^{(1)}} = { {\tilde u} \choose {\tilde d} } \ , \
{\tilde q^{(2)}} = { {\tilde c} \choose {\tilde s} } \ , \
{\tilde q^{(3)}} = { {\tilde t} \choose {\tilde b} } \ ,
\label{2HDnot2}
\end{equation}
\begin{equation}
\langle {\Phi}^{(1)} \rangle_0 = 
\frac{{\mathrm{e}}^{{\mathrm{i}}\eta_1}}{\sqrt{2}} 
{0 \choose v_1} \ ,
\qquad
\langle {\Phi}^{(2)} \rangle_0 = \
\frac{{\mathrm{e}}^{{\mathrm{i}}\eta_2}}{\sqrt{2}} 
{0 \choose v_2} \ ,
\qquad v_1^2+v_2^2 = v^2 \ .
\label{2HDnot3}
\end{equation}
In (\ref{2HDnot3}), $v$ [$\equiv v(E)$] is the usual VEV 
needed for the electroweak symmetry breaking, \ie{},
$v(E_{\rm{ew}})\!\approx\!246$ GeV. 
The phase difference $\eta\!\equiv\!\eta_2\!-\!\eta_1$ 
between the two VEV's in (\ref{2HDnot3}) may be nonzero; 
it represents CP violation originating at low energies 
from the scalar 2HD sector (cf.~\cite{Gunionetal}).
The leptonic sector will be ignored throughout.

We note that the popular ``type I'' and ``type II'' models
are special cases (subsets) of this framework, 
with some of the Yukawa matrices being exactly zero:
${\tilde U}^{(1)}\!=\!{\tilde D}^{(1)}\!=\!0$ [2HDM(I)];
${\tilde U}^{(1)}\!=\!{\tilde D}^{(2)}\!=\!0$ [2HDM(II)].
In these two special models, suggested by Glashow and Weinberg 
\cite{GlashowWeinberg}, the flavor-changing neutral (FCN)
Yukawa coupling parameters are exactly
zero. This is so because one of the two nonzero Yukawa matrices 
is proportional to the mass matrix of the up-type quarks,
and the other to the mass matrix of the down-type quarks.
Since flavors refer to the physical quarks, and the quark mass 
matrices in the physical (mass) basis are diagonal by
definition, the FCN Yukawa coupling parameters 
(i.e., off-diagonal elements) are zero. 
Moreover, this is true at any energy of probes 
(cutoff energy) $E^{\prime}$. Stated otherwise, when the
original cutoff $E$ is changed to $E^{\prime}$, no 
loop-induced ($\ln E^{\prime}$ cutoff-dependent) FCN
Yukawa coupling parameters appear, i.e., 
the original form of the Lagrangian is
preserved under the change of the cutoff.
This can be formulated also in terms of 
explicit ${\mathrm{U(1)}}$-type
family symmetries governing the Yukawa part of the
Lagrangian density in the 2HDM(I) and 2HDM(II). 
These symmetries ensure that, in the course of
the change of the cutoff (i.e., evolution energy or the
energy of probes), the original form of the Yukawa
Lagrangian density is preserved.
In the 2HDM(II) the symmetry tranformation is:
${\tilde d}^{(j)}_{R}\!\rightarrow\!
{\mathrm{e}}^{{\mathrm{i}} \alpha} {\tilde d}^{(j)}_{R}$, 
${\Phi}^{(1)}\!\rightarrow\! 
{\mathrm{e}}^{-{\mathrm{i}} \alpha} {\Phi}^{(1)}$ 
($j\!=\!1,2,3$), the other fields remaining unchanged; 
in the 2HDM(I): ${\tilde d}^{(j)}_R\!\rightarrow\!
{\mathrm{e}}^{{\mathrm{i}} \alpha} {\tilde d}^{(j)}_R$,   
${\tilde u}^{(j)}_R\!\rightarrow\!
{\mathrm{e}}^{-{\mathrm{i}} \alpha} {\tilde u}^{(j)}_R$, 
${\Phi}^{(2)}\!\rightarrow\!
{\mathrm{e}}^{-{\mathrm{i}} \alpha} {\Phi}^{(2)}$.

In contrast to the 2HDM(I) and 2HDM(II), the G2HDM 
has no explicit family symmetry 
enforcing the complete suppression
of the FCN Yukawa coupling parameters. There are at least
two consequences of this fact:
\begin{enumerate}
\item 
The FCN Yukawa parameters in the G2HDM are in general nonzero.
At low energies of probes ($E\!\sim\!E_{\rm{ew}}$), 
those FCN Yukawa parameters which don't involve the top
quark must be given quite small values (not necessarily
zero) for phenomenological viability of the model.
\item
Even if the FCN Yukawa parameters are zero at
some low energy of probes, they become in general
nonzero at higher energies. If some FCN Yukawa
parameters are small (but nonzero) at small energies,
there exists in principle the possibility that
they increase by a large factor, or even by orders of
magnitude, when the energy of probes increases (but
remains a safe distance away from the Landau pole)
-- due to the absence of an explicit 
protective family symmetry.
\end{enumerate}
The Lagrangian density (\ref{2HD30})\footnote{
Throughout this Section we omit, for simpler notation,
reference to the evolution (cutoff) energy $E$ at the quark
fields, at the scalar fields and their VEV's, and at the Yukawa
coupling parameters.}  
can be written in a form more convenient for consideration of
the FCN Yukawa coupling parameters, by redefining the scalar
isodoublets in the following way:
\begin{eqnarray}
{\Phi}^{\prime(1)}&  =&  (\cos \beta) {\Phi}^{(1)} +
         (\sin \beta){\mathrm{e}}^{-{\mathrm{i}} 
{\eta}} {\Phi}^{(2)} \ , 
\nonumber\\
{\Phi}^{\prime(2)} & =&  - (\sin \beta) {\Phi}^{(1)} +
         (\cos \beta){\mathrm{e}}^{-{\mathrm{i}} 
{\eta}} {\Phi}^{(2)} \ ,
\label{redefPhi}
\\
\mbox{where: } \
\eta&=&\eta_2-\eta_1 \ ; \quad
\tan \beta = \frac{v_2}{v_1} \ \Rightarrow \
\cos \beta = \frac{v_1}{v} \ , \quad \sin \beta = \frac{v_2}{v} \ .
\label{xiVEV}
\end{eqnarray}
Therefore, the VEV's of the redefined scalar isodoublets are
\begin{equation}
{\mathrm{e}}^{-{\mathrm{i}} {\eta}_1} 
\langle {\Phi}^{\prime(1)} \rangle_0 = \frac{1}{\sqrt{2}}
{0 \choose v} \ , \qquad
\langle {\Phi}^{\prime(2)} \rangle_0 = \frac{1}{\sqrt{2}}
{0 \choose 0} \ .
\label{newVEVs}
\end{equation}
The isodoublet ${\Phi}^{\prime(1)}$ is therefore responsible
for the masses of the quarks, and couplings of
${\Phi}^{\prime(2)}$ to the quarks lead to the FCN Yukawa
couplings, as will be seen below.
The original Yukawa Lagrangian density (\ref{2HD30}) of 
the G2HDM can then be rewritten 
in terms of these redefined scalar fields as
\begin{eqnarray}
{\cal L}^{(E)}_{ \rm G2HDM } & = & 
- \sum_{i,j=1}^3 {\Big \lbrace}
{\tilde G}^{(D)}_{ij}( {\bar {\tilde q}}^{(i)}_L {\Phi}^{\prime (1)} )
{\tilde d}^{(j)}_{R} +
{\tilde G}^{(U)}_{ij}( {\bar {\tilde q}}^{(i)}_L {\tilde \Phi}^{\prime(1)} )
{\tilde u}^{(j)}_{R} + {\mathrm{h.c. }} {\Big \rbrace}
\nonumber\\
&&
- \sum_{i,j=1}^3 {\Big \lbrace}
{\tilde D}_{ij}( {\bar {\tilde q}}^{(i)}_L {\Phi}^{\prime (2)} )
{\tilde d}^{(j)}_{R} +
{\tilde U}_{ij}( {\bar {\tilde q}}^{(i)}_L {\tilde \Phi}^{\prime(2)} )
{\tilde u}^{(j)}_{R} + {\mathrm{h.c. }} {\Big \rbrace} \ ,
\label{Lnew}
\end{eqnarray}
where the Yukawa matrices ${\tilde G}^{(U)}$ and ${\tilde G}^{(D)}$
are rescaled mass matrices, and ${\tilde U}$ and ${\tilde D}$
the corresponding ``complementary'' Yukawa matrices, in an
(arbitrary) $\rm{SU(2)_L}$-basis (weak basis)
\begin{eqnarray}
{\tilde G}^{(U)}& = &{\sqrt{2}} {\tilde M}^{(U)}/v =
(\cos \beta) {\tilde U}^{(1)} + 
(\sin \beta) {\mathrm{e}}^{-{\mathrm{i}} {\eta}} {\tilde U}^{(2)} \ ,
\nonumber\\
{\tilde G}^{(D)}& =& {\sqrt{2}} {\tilde M}^{(D)}/v =
(\cos \beta) {\tilde D}^{(1)} + 
(\sin \beta) {\mathrm{e}}^{+{\mathrm{i}} {\eta}} {\tilde D}^{(2)} \ ;
\label{Gs}
\\
{\tilde U} & = & - (\sin \beta) {\tilde U}^{(1)} +
(\cos \beta) {\mathrm{e}}^{-{\mathrm{i}}{\eta}} {\tilde U}^{(2)} \ ,
\nonumber\\
{\tilde D} & = & - (\sin \beta) {\tilde D}^{(1)} +
(\cos \beta) {\mathrm{e}}^{+{\mathrm{i}}{\eta}} {\tilde D}^{(2)} \ .
\label{UDs}
\end{eqnarray}
By a biunitary transformation involving unitary matrices
$V_L^U$, $V_R^U$, $V_L^D$ and $V_R^D$, the Yukawa
parameters can be expressed in the mass basis of the quarks, 
where the (rescaled) mass matrices $G^{(U)}$ and $G^{(D)}$ are
diagonal and real
\begin{eqnarray}
\!\!\!\!\!\!\!
G^{(U)} = \frac{\sqrt{2}}{v} M^{(U)}& = &
V_L^{U} {\tilde G}^{(U)} V_R^{U\dagger} 
\ \ \left[ M_{ij}^{(U)}= {\delta}_{ij} m_i^{(u)} \right] \ ;
\ \ U  =  V_L^{U} {\tilde U} V_R^{U\dagger} \ ;
\label{GUUmass}
\\
\!\!\!\!\!\!\!
G^{(D)} = \frac{\sqrt{2}}{v} M^{(D)}& =&
V_L^{D} {\tilde G}^{(D)} V_R^{D\dagger} 
\ \ \left[ M_{ij}^{(D)}= {\delta}_{ij} m_i^{(d)} \right] \ ;
\ \ D  =  V_L^{D} {\tilde D} V_R^{D\dagger} \ ;
\label{GDDmass}
\end{eqnarray}
\begin{equation}
u_L = V_L^{U} {\tilde u}_L \ , \quad u_R = V_R^U {\tilde u}_R \ , \quad
d_L = V_L^{D} {\tilde d}_L \ , \quad d_R = V_R^D {\tilde d}_R \ .
\label{qmass}
\end{equation}
The absence of tildes above the Yukawa coupling parameters
and above the quark fields means that these quantities are in the
quark mass basis (at a given evolution energy $E$). 
Lagrangian density (\ref{Lnew}) can be written now in the
quark mass basis. The ``neutral current'' part of the Lagrangian 
density in the quark mass basis is
\begin{eqnarray}
{\cal L}^{ (E){\rm n.c.} }_{ \rm{G2HDM} } &=&
- \frac{1}{\sqrt{2}} \sum_{i=1}^3  
{\Big \lbrace}
G^{(D)}_{ii} {\bar d}^{(i)}_L d^{(i)}_{R} 
( {\phi}^{\prime(1)}_3\!+\!{\mathrm{i}} {\phi}^{\prime(1)}_4 ) +
\nonumber\\ 
&& + G^{(U)}_{ii} {\bar u}^{(i)}_L u^{(i)}_{R}
( {\phi}^{\prime(1)}_3\!-\!{\mathrm{i}} {\phi}^{\prime(1)}_4 )  
+ {\rm{h.c. }} {\Big \rbrace}
\nonumber\\
&&
- \frac{1}{\sqrt{2}} \sum_{i,j=1}^3 {\Big \lbrace}
D_{ij} {\bar d}^{(i)}_L d^{(j)}_{R} 
( {\phi}^{\prime(2)}_3\!+\!{\mathrm{i}} {\phi}^{\prime(2)}_4 ) +  
\nonumber\\
&& + U_{ij} {\bar u}^{(i)}_L u^{(j)}_{R}
( {\phi}^{\prime(2)}_3\!-\!{\mathrm{i}} {\phi}^{\prime(2)}_4 )  
 + {\mathrm{h.c. }} {\Big \rbrace} \ .
\label{Lmassn}
\end{eqnarray}
On the other hand, the ``charged current'' part of the 
Lagrangian density in the quark mass basis is
\begin{eqnarray}
{\cal L}^{ (E) {\rm c.c.} }_{ \rm{G2HDM} } & =&  
- \frac{1}{\sqrt{2}} \sum_{i,j=1}^3  
{\Big \lbrace}
(V G^{(D)})_{ij} {\bar u}^{(i)}_L d^{(j)}_{R} 
( {\phi}^{\prime(1)}_1\!+\!{\mathrm{i}} {\phi}^{\prime(1)}_2 )  - 
\nonumber\\
&& - (V^{\dagger} G^{(U)})_{ij} {\bar d}^{(i)}_L u^{(j)}_{R}
( {\phi}^{\prime(1)}_1\!-\!{\mathrm{i}} {\phi}^{\prime(1)}_2 )  
+ {\mathrm{h.c. }} {\Big \rbrace}
\nonumber\\
&&
- \frac{1}{\sqrt{2}} \sum_{i,j=1}^3 {\Big \lbrace}
(V D)_{ij} {\bar u}^{(i)}_L d^{(j)}_{R} 
( {\phi}^{\prime(2)}_1\!+\!{\mathrm{i}} {\phi}^{\prime(2)}_2 ) -
\nonumber\\ 
&&-(V^{\dagger} U)_{ij} {\bar d}^{(i)}_L u^{(j)}_{R}
( {\phi}^{\prime(2)}_1\!-\!{\mathrm{i}} {\phi}^{\prime(2)}_2 )  
 + {\mathrm{h.c. }} {\Big \rbrace} \ .
\label{Lmassch}
\end{eqnarray}
Here, we denoted by $V$ the Cabibbo-Kobayashi-Maskawa (CKM) matrix
\begin{equation}
V \equiv V_{\mathrm{CKM}} = V_L^U V_L^{D\dagger} \ .
\label{CKM}
\end{equation}
We see from (\ref{Lmassn}) that the $U$ and $D$ matrices,
as defined by (\ref{UDs}) and (\ref{GUUmass})-(\ref{GDDmass})
through the original Yukawa matrices ${\tilde U}^{(j)}$ and
${\tilde D}^{(j)}$ of the G2HDM Lagrangian density
(\ref{2HD30}), allow the model to possess 
in general scalar-mediated FCNC's. Namely, in the quark mass 
basis only the (rescaled) 
quark mass matrices $G^{(U)}$ and $G^{(D)}$ 
of (\ref{GUUmass})-(\ref{GDDmass}) [cf.~also (\ref{Gs})]
are diagonal, but the matrices $U$ and $D$ in this
general framework are in general not diagonal. The off-diagonal
elements of the matrices $U$ and $D$ are the FCN Yukawa coupling
parameters
\begin{eqnarray}
\lefteqn{ \!\!\!\!\!\!\!
{\cal L}^{ (E) {\rm FCN} }_{\rm{G2HDM}}  =
- \frac{1}{ \sqrt{2} } 
\sum_{
\begin{array}{c}
\vspace{-7.mm}
{\scriptstyle i,j = 1} \\[-4.mm]
{\scriptstyle i \not= j}
\end{array}}^3
{\Big \lbrace}
D_{ij} {\bar d}^{(i)}_L d^{(j)}_{R} 
( {\phi}^{\prime(2)}_3\!+\!{\mathrm{i}} {\phi}^{\prime(2)}_4 ) +  
(D^{\dagger})_{ij} {\bar d}^{(i)}_R d^{(j)}_L
( {\phi}^{\prime(2)}_3\!-\!{\mathrm{i}} {\phi}^{\prime(2)}_4 ) 
{\Big \rbrace} }
\nonumber\\
&&
- \frac{1}{ \sqrt{2} } 
\sum_{
\begin{array}{c}
\vspace{-7.mm}
{\scriptstyle i,j = 1} \\[-4.mm]
{\scriptstyle i \not= j}
\end{array}}^3
{\Big \lbrace}
U_{ij} {\bar u}^{(i)}_L u^{(j)}_{R}
( {\phi}^{\prime(2)}_3\!-\!{\mathrm{i}} {\phi}^{\prime(2)}_4 ) + 
(U^{\dagger})_{ij} {\bar u}^{(i)}_R u^{(j)}_L
( {\phi}^{\prime(2)}_3\!+\!{\mathrm{i}} {\phi}^{\prime(2)}_4 )  
{\Big \rbrace} \ .
\label{FCNCs}
\end{eqnarray}
It should be noted that the original four Yukawa matrices
${\tilde U}^{(j)}$ and ${\tilde D}^{(j)}$ ($j=1,2$)
in an $\mathrm{SU(2)_L}$-basis
are already somewhat constrained by the requirement that (at low
energy) the squares of\footnote 
{Strictly speaking, the following
``squares'': $M^{(U)} M^{(U)\dag}$ and $M^{(D)} M^{(D)\dag}$.} 
$M^{(U)}$ and $M^{(D)}$ are
diagonalized by unitary transformations
involving such unitary matrices $V_L^U$ and $V_L^D$,
respectively, which are related
to each other by $V_L^U V_L^{D\dag} = V$. Here, $V$ is the
CKM matrix which is, for any specific chosen
phase convention, more or less known at low energies.

In order to have at low evolution energies 
($E\!\sim\!E_{\rm{ew}}$) a phenomenologically viable
suppression of the scalar-mediated FCNC's, the
authors Cheng, Sher and Yuan (CSY) 
\cite{CSY} basically argued that
the elements of the $U$ and $D$ matrices (in the quark 
mass basis and at low evolution energies $E$) should have the form:
\begin{equation}
U_{ij}(E) = {\xi}^{(u)}_{ij} \frac{\sqrt{2}}{v}
\sqrt{m_i^{(u)} m_j^{(u)}} \ , \qquad
D_{ij}(E) = {\xi}^{(d)}_{ij} \frac{\sqrt{2}}{v}
\sqrt{m_i^{(d)} m_j^{(d)}} \ , 
\label{FCNCcon1}
\end{equation}
\begin{equation}
\mbox{where} \qquad
{\xi}^{(u)}_{ij}, {\xi}^{(d)}_{ij} \sim 1 \quad {\rm{for }} \ 
E \sim E_{\mathrm{ew}} \ .
\label{FCNCcon2}
\end{equation}
This form is in general phenomenologically acceptable. It is
strongly motivated by the actual mass hierarchies of the quarks. 
At least for the diagonal elements, it is suggested
by the requirement that (at a given {\em low\/} energy
$\sim\!E_{\rm{ew}}$) there be no fine-tuning on the
right of Eqs.~(\ref{Gs})-(\ref{UDs})
when these equations are written in the quark mass
basis (\ie{}, no tildes over the matrices).
For definiteness, consider the up-type sector. 
The diagonal elements $U^{(i)}_{jj}$ are in general 
$\sim\!m_j^{(u)}/v$ unless fine-tuning is 
involved on the right of (\ref{Gs}). Consequently, 
also $U_{jj}\!\sim\!m_j^{(u)}/v$ unless fine-tuning is 
involved on the right of (\ref{UDs}).
This consideration further suggests (but not necessarily implies)
that the off-diagonal elements $U^{(i)}_{jk}$ 
and $U_{jk}$ have values between those 
of the corresponding diagonal
elements $U^{(i)}_{jj}\!\sim\!m_j^{(u)}/v$ and 
$U^{(i)}_{kk}\!\sim\!m_k^{(u)}/v$,
for example roughly the geometrical mean of those,
leading thus to the CSY ansatz 
(\ref{FCNCcon1})-(\ref{FCNCcon2}).\footnote{
For the complete suppression of FCN Yukawa couplings
$U_{jk}\!=\!0$ (for $j\!\not=\!k$) we would
then need fine-tuning on the right of (\ref{UDs}).}
Therefore, this (CSY) form is
considered to be reasonably natural. 

{}From the CSY ansatz (\ref{FCNCcon1})-(\ref{FCNCcon2})
we see that the FCN Yukawa vertices
involving the heavy top quark are the only ones that are not
strongly suppressed (at low evolution energies).
As mentioned in the Introduction, scalar-exchange-mediated
FCNC processes involving the top quark vertices (not loops
with top quarks) are not constrained
by present experiments. Later in Section 4 we will
use low energy conditions (\ref{FCNCcon1})-(\ref{FCNCcon2}) 
for a numerical investigation of the RGE flow of the FCN Yukawa 
coupling parameters.

\section{One-loop RGE's in a convenient parametrization}

In the Appendix we outlined a derivation of
the relevant set of one-loop RGE's for the scalar fields 
and their VEV's (\ref{RGEphi1Y})-(\ref{RGEVEVs}), 
for the quark fields (\ref{ansq}), (\ref{fLres})-(\ref{fdRres}),
and for the Yukawa matrices 
${\tilde U}^{(k)}$ and ${\tilde D}^{(k)}$ 
(\ref{RGEUk})-(\ref{RGEDk}). One of the reasons
for performing an independent derivation is that we consider
the method of finite cutoffs~\cite{Lepage},
which was used in the derivation, as physically very
intuitive. This contrasts with other methods often 
applied in the literature, which are, however,
usually mathematically more efficient at two-loop
and higher-loop levels. Another reason is 
that there is a certain
disagreement between the results on one-loop beta functions
derived for a general (semi)simple gauge group $G$
in various parts of literature -- see comparisons and the
discussion toward the end of this Section. 

We can rewrite all the RGE's derived in the Appendix,
now in a
a more convenient set of parameters. These are:
the VEV parameters $v \equiv \sqrt{v_1^2 + v_2^2}$, 
$\tan \beta \equiv v_2/v_1$ and 
$\eta \equiv \eta_2-\eta_1$
[cf.~(\ref{2HDnot3})], and matrices ${\tilde G}^{(U)}$,
${\tilde G}^ {(D)}$, ${\tilde U}$ and ${\tilde D}$
[cf.~(\ref{Gs}), (\ref{UDs})] -- this representation
is more convenient for discerning
the running of the FCN Yukawa coupling parameters.
Applying lengthy, but straightforward, algebra to the
hitherto obtained RGE's then results in the following 
RGE's in terms of the mentioned set of parameters:
\begin{eqnarray}
16 \pi^2 \frac{d (v^2)}{d \ln E} & = &
- 2 N_{\mathrm{c}} {\mathrm{Tr}} \left[
{\tilde G}^{(U)} {\tilde G}^ {(U)\dagger} 
+ {\tilde G}^{(D)} {\tilde G}^ {(D)\dagger} \right] v^2
+ \left[ \frac{3}{2} g_1^2 + \frac{9}{2} g_2^2 \right] v^2 ,
\label{RGEv2}
\end{eqnarray}
\begin{eqnarray}
16 \pi^2 \frac{d (\tan \beta) }{d \ln E} & = &
- \frac{N_{\mathrm{c}}}{2 \cos^2 \beta} 
{\mathrm{Tr}} \left[
{\tilde G}^{(U)} {\tilde U}^{\dagger} +
{\tilde U} {\tilde G}^{(U)\dagger}
+ {\tilde G}^{(D)} {\tilde D}^ {\dagger} 
+ {\tilde D} {\tilde G}^ {(D)\dagger} \right] ,
\label{RGEtbeta}
\end{eqnarray}
\begin{eqnarray}
16 \pi^2 \frac{ d (\eta) }{d \ln E} & = &
\frac{N_{\mathrm{c}}}{{\mathrm{i}} \sin (2 \beta)}
{\mathrm{Tr}} \left[ 
{\tilde G}^{(U)} {\tilde U}^{\dagger}
- {\tilde U} {\tilde G}^ {(U)\dagger} 
- {\tilde G}^{(D)} {\tilde D}^{\dagger}
+ {\tilde D} {\tilde G}^{(D)\dagger} \right] ,
\label{RGExi}
\end{eqnarray}
\begin{eqnarray}
\lefteqn{
16 \pi^2 \frac{d}{d \ln E} ({\tilde U}) =
N_{\mathrm{c}} {\Bigg \{}
2 {\mathrm{Tr}} \left[ {\tilde U} {\tilde G}^{(U)\dagger}
+ {\tilde G}^{(D)} {\tilde D}^{\dagger} \right] {\tilde G}^ {(U)}
+ {\mathrm{Tr}} \left[ {\tilde U} {\tilde U}^ {\dagger} +
{\tilde D} {\tilde D}^ {\dagger} \right] {\tilde U} {\Bigg \}}
 }
\nonumber\\
&&
+ \frac{1}{2} N_{\rm c} (\cot \beta) {\tilde U}
{\mathrm{Tr}} \left[ - {\tilde G}^{(U)} {\tilde U}^{\dagger}
+ {\tilde U} {\tilde G}^{(U)\dagger} 
+ {\tilde G}^{(D)} {\tilde D}^{\dagger}
- {\tilde D} {\tilde G}^{(D)\dagger} \right] 
\nonumber\\
&& + {\Bigg \{} \frac{1}{2} \left[ {\tilde U} {\tilde U}^{\dagger}
+ {\tilde D} {\tilde D}^{\dagger} + {\tilde G}^{(U)} {\tilde G}^ {(U)\dagger}
+ {\tilde G}^{(D)} {\tilde G}^ {(D)\dagger} \right] {\tilde U} +
{\tilde U} \left[ {\tilde U}^{\dagger} {\tilde U} +
{\tilde G}^{(U)\dagger} {\tilde G}^{(U)} \right] 
\nonumber\\
&&
 - 2 {\tilde D} {\tilde D}^{\dagger} {\tilde U}
- 2 {\tilde G}^{(D)} {\tilde D}^ {\dagger} {\tilde G}^ {(U)} 
- A_U {\tilde U} {\Bigg \}} \ ,
\label{RGEU}
\end{eqnarray}
\begin{eqnarray}
\lefteqn{
16 \pi^2 \frac{d}{d \ln E}({\tilde D}) =
N_{\mathrm{c}} {\Bigg \{}
2 {\mathrm{Tr}} \left[ {\tilde D} {\tilde G}^{(D)\dagger}
+ {\tilde G}^{(U)} {\tilde U}^{\dagger} \right] {\tilde G}^ {(D)}
+ {\mathrm{Tr}} \left[ {\tilde U} {\tilde U}^ {\dagger} +
{\tilde D} {\tilde D}^ {\dagger} \right] {\tilde D} {\Bigg \}}
  }
\nonumber\\
&&
+ \frac{1}{2} N_{\rm c} (\cot \beta) {\tilde D}
{\mathrm{Tr}} \left[ - {\tilde G}^{(D)} {\tilde D}^{\dagger}
+ {\tilde D} {\tilde G}^{(D)\dagger} 
+ {\tilde G}^{(U)} {\tilde U}^{\dagger}
- {\tilde U} {\tilde G}^{(U)\dagger} \right] 
\nonumber\\
&& + {\Bigg \{} \frac{1}{2} \left[ {\tilde U} {\tilde U}^{\dagger}
+ {\tilde D} {\tilde D}^{\dagger} + {\tilde G}^{(U)} {\tilde G}^ {(U)\dagger}
+ {\tilde G}^{(D)} {\tilde G}^ {(D)\dagger} \right] {\tilde D} +
{\tilde D} \left[ {\tilde D}^{\dagger} {\tilde D} +
{\tilde G}^{(D)\dagger} {\tilde G}^{(D)} \right] 
\nonumber\\
&&
 - 2 {\tilde U} {\tilde U}^{\dagger} {\tilde D}
- 2 {\tilde G}^{(U)} {\tilde U}^ {\dagger} {\tilde G}^ {(D)} 
- A_D {\tilde D} {\Bigg \}} \ ,
\label{RGED}
\end{eqnarray}
\begin{eqnarray}
\lefteqn{
16 \pi^2 \frac{d}{d \ln E} \left( {\tilde G}^{(U)}\right) =
N_{\mathrm{c}} 
 {\mathrm{Tr}} \left[ {\tilde G}^{(U)} 
{\tilde G}^ {(U)\dagger} +
{\tilde G}^{(D)} {\tilde G}^ {(D)\dagger} 
\right] {\tilde G}^{(U)}
 }
\nonumber\\
&&
+ \frac{1}{2} N_{\mathrm{c}} (\tan \beta) {\tilde G}^{(U)}
{\mathrm{Tr}} \left[ 
-{\tilde G}^{(U)} {\tilde U}^{\dagger}
+{\tilde U} {\tilde G}^{(U)\dagger} 
+ {\tilde G}^{(D)} {\tilde D}^{\dagger}
- {\tilde D} {\tilde G}^{(D)\dagger} 
\right] 
\nonumber\\
&& 
+ \frac{1}{2} \left[ {\tilde U} {\tilde U}^{\dagger}
+ {\tilde D} {\tilde D}^{\dagger} 
+ {\tilde G}^{(U)} {\tilde G}^ {(U)\dagger}
+ {\tilde G}^{(D)} {\tilde G}^ {(D)\dagger} 
\right] {\tilde G}^{(U)} 
\nonumber\\
&& 
\!\!\!\!\!\!\!
+ {\tilde G}^ {(U)} \left[ {\tilde U}^{\dagger} {\tilde U} 
+ {\tilde G}^{(U)\dagger} {\tilde G}^{(U)} \right] 
 - 2 {\tilde D} {\tilde G}^{(D)\dagger} {\tilde U}
- 2 {\tilde G}^{(D)} {\tilde G}^ {(D)\dagger} {\tilde G}^ {(U)} 
- A_U {\tilde G}^{(U)} ,
\label{RGEGU}
\end{eqnarray}
\begin{eqnarray}
\lefteqn{
16 \pi^2 \frac{d}{d \ln E} \left( {\tilde G}^{(D)} \right) =
N_{\mathrm{c}} 
 {\mathrm{Tr}} \left[ {\tilde G}^{(U)} {\tilde G}^ {(U)\dagger} +
{\tilde G}^{(D)} {\tilde G}^ {(D)\dagger} \right] {\tilde G}^{(D)}
  }
\nonumber\\
&&
+ \frac{1}{2} N_{\mathrm{c}} (\tan \beta) {\tilde G}^{(D)}
{\mathrm{Tr}} \left[ 
-{\tilde G}^{(D)} {\tilde D}^{\dagger}
+{\tilde D} {\tilde G}^{(D)\dagger} 
+ {\tilde G}^{(U)} {\tilde U}^{\dagger}
- {\tilde U} {\tilde G}^{(U)\dagger} 
\right] 
\nonumber\\
&& 
+ \frac{1}{2} \left[ {\tilde U} {\tilde U}^{\dagger}
+ {\tilde D} {\tilde D}^{\dagger} 
+ {\tilde G}^{(U)} {\tilde G}^ {(U)\dagger}
+ {\tilde G}^{(D)} {\tilde G}^ {(D)\dagger} \right] {\tilde G}^{(D)} 
\nonumber\\
&& 
\!\!\!\!\!\!\!
+ {\tilde G}^ {(D)} \left[ {\tilde D}^{\dagger} {\tilde D} 
  + {\tilde G}^{(D)\dagger} {\tilde G}^{(D)} \right] 
 - 2 {\tilde U} {\tilde G}^{(U)\dagger} {\tilde D}
- 2 {\tilde G}^{(U)} {\tilde G}^ {(U)\dagger} {\tilde G}^ {(D)} 
- A_D {\tilde G}^{(D)} .
\label{RGEGD}
\end{eqnarray}
Equation (\ref{RGEv2}) is in the Landau gauge, while the
other RGE's (\ref{RGEtbeta})-(\ref{RGEGD}) are gauge independent.

For a general (semi)simple gauge group $G$, 
RGE's for various parameters have been derived
at one-loop level
\cite{ChengEichtenLi}-\cite{Vaughn}, and at
two-loop level \cite{MV}-\cite{VD}. First we should note
that these groups of authors are using conventions
which, particularly as to the fermionic sector,
differ from each other. 

Cheng, Eichten and Li \cite{ChengEichtenLi}
were using the usual four-component Dirac spinors for quarks.
While they allowed an arbitrary number of (real) scalar
degrees of freedom, their one-loop RGE's for the
Yukawa coupling parameters are directly applicable
only when these parameters are real (noncomplex).

The one-loop RGE results of Vaughn \cite{Vaughn} and 
one- and two-loop results of Machacek and Vaughn \cite{MV}
were written for the general case of complex Yukawa
coupling parameters. 
Their scalar fields ${\phi}_a$ were real,
and for fermions (quarks) they were
using two-component spinor fields ${\psi}_j$ as defined 
by Sikivie and G\"ursey \cite{SG}. 
The Yukawa Lagrangian density was written in 
\cite{Vaughn}-\cite{MV} in the form
\begin{equation}
{\cal {L}}_{\rm Y.} = -  {\bf Y}^a_{jk} {\psi}_j^T 
{\mathrm {i}} {\sigma}^2
{\psi}_k {\phi}_a + {\rm h.c.} \ ,
\label{VaMa}
\end{equation}
where ${\phi}_a$ are the real scalar degrees of freedom,
and for fermions (quarks) the two-component left-handed 
spinor fields ${\psi}_j$ are used as defined in Ref.~\cite{SG};
${\sigma}^2$ is the second Pauli matrix. The double sum
over spinor degrees of freedom in (\ref{VaMa}) is running over
$j,k = 1, \ldots, 2n$, where $n$ is the number of
fermion (quark) flavors. The usual four-component
Dirac spinor fields ${\Psi}^{(j)}$ 
in the chiral basis (i.e, the basis
of Ref.~\cite{Peskin}) are then, according to Sikivie
and G\"ursey \cite{SG}\footnote{
Sikivie and G\"ursey \cite{SG} use the notation $F_A$, $F_B$ for
${\psi}_j$, ${\psi}_k$, respectively, and
$(- \Gamma^{i})$ and ${\phi}_i$ for ${\bf Y}^a$, ${\phi}_a$,
respectively.}
\begin{equation}
{\Psi}^{(j)} = \left[
\begin{array}{c}
{\psi}_j \\
- {\mathrm {i}} {\sigma}^2 {\psi}^{\dagger T}_{j+n}
\end{array}
\right] \qquad  (j=1,\ldots, n) \ .
\label{SikGuer}
\end{equation}
If we carefully rewrite our RGE's
(from the Appendix) for the scalar and quark fields, 
and for the Yukawa parameters,
in terms of the spinor fields of \cite{SG} and
of real scalar field components, we notice several
differences when compared with one-loop results
of \cite{Vaughn}-\cite{MV}. We can deduce from
our RGE's for the Yukawa coupling parameters
that RG Equation (3.4) of \cite{MV} (second entry),
or equivalently, RG Eqs.~(2.2)-(2.3) of \cite{Vaughn},
for ${\bf Y}^a$ Yukawa matrices in their language
should read:
\begin{eqnarray}
\lefteqn{
\left( 4 \pi \right)^2 \frac{ d {\bf Y}^a}{d \ln E}{\Bigg |}_{\rm 1-l.}
 \equiv  (4 \pi )^2 {\bf \beta}^{a} {\Big |}_{\rm 1-l.} =
2 \left[ {\bf Y}^b {\bf Y}^{b \dagger} {\bf Y}^a\!+\!
{\bf Y}^a {\bf Y}^{b \dagger} {\bf Y}^b \right]\!+\!
8 {\bf Y}^b {\bf Y}^{a \dagger} {\bf Y}^b\!+ }
\nonumber\\
&&
\,\,\,\,\,\,\,\,\,\,
+ 4 {\kappa} {\bf Y}^b {\rm Tr} \left( {\bf Y}^{b \dagger} {\bf Y}^a
\!+\! {\bf Y}^{a \dagger} {\bf Y}^b \right) - 3 g^2
\left\{ {\bf C}_2(F), {\bf Y}^a \right\} ,
\label{MV3.4}
\end{eqnarray}
where ${\kappa}\!=\!1/2$. Stated otherwise, the cubic
Yukawa terms on the right of this RGE are effectively
those given in \cite{Vaughn}-\cite{MV}, but multiplied\footnote{
If Machacek and Vaughn had introduced in the Lagrangian
density an additional factor of (1/2)
in front of the sum (\ref{VaMa}) ( -- but they didn't), 
the factor 4 in the cubic Yukawa terms on the right
of (\ref{MV3.4}) wouldn't have occurred.}
by factor $4$, and the trace there is replaced now by
its real (symmetric) part. Similar differences arise when comparing
our RGE's for scalar and quark fields with those of
\cite{MV}. Instead of Eq.~(3.7) of \cite{MV} (first entry),
we get
\begin{equation}
(4 \pi)^2 \gamma^s_{ab}|_{\rm 1-l.} =
4 {\kappa} {\rm Tr} \left( {\bf Y}^a {\bf Y}^{b \dagger}
+ {\bf Y}^{a \dagger} {\bf Y}^b \right) - g^2 (2\!+\!\alpha)
{\bf C}_2(S) \delta_{ab} \ ,
\label{MV3.7}
\end{equation}
where again ${\kappa}\!=\!1/2$ and $\gamma^s$ is defined via:
$d \phi_a/d \ln E\!=\!-\gamma^s_{ab} \phi_b$. Here,
$\phi_a$ are the real scalar fields, and $\alpha$ in
(\ref{MV3.7}) is the gauge parameter ($\alpha\!\equiv\!1-\xi\!=\!1$ in the
Landau gauge). Instead of Eq.~(4.5) of \cite{MV} (first entry),
we get
\begin{equation} 
(4 \pi)^2 \gamma^F|_{\rm 1-l.} = 2 {\bf Y}^{a \dagger} {\bf Y}^a
+ g^2 {\bf C}_2(F) (1\!-\!\alpha) \ ,
\label{MV4.5}
\end{equation}
where $\gamma^F$ is defined via: 
$d \psi_j/d \ln E\!=\!-\gamma^F_{ji} \psi_i$. Here,
$\psi_i$ are left-handed two-component spinors as defined
in \cite{SG}.

The authors of \cite{JO}, on the other hand, worked with Majorana
fermions, using background field method. 
The Dirac fermions can then be expressed as sums of two
Majorana fermions. Their one-loop beta functions for the
Yukawa coupling parameters of the (real) scalars with
the (Majorana) fermions, can be re-expressed 
in the notation with left-handed two-component
spinors $\psi_j$ as introduced by \cite{SG} and used by
\cite{Vaughn}-\cite{MV}. After a somewhat lengthy algebra,
it can be shown that the one-loop results of \cite{JO} 
lead precisely to formula (\ref{MV3.4}). Therefore, we finally conclude 
that our one-loop RGE formulas for the Yukawa coupling 
parameters, derived in the Appendix and rewritten in
Eq.~(\ref{MV3.4}) in the language of \cite{SG}, 
{\em are not in agreement with\/} those of Vaughn \cite{Vaughn}
and of Machacek and Vaughn \cite{MV},
and {\em are in agreement with\/} 
the results of Jack and Osborn \cite{JO}.
Moreover, the latter authors emphasize that their RGE results agree
with those of van Damme \cite{VD}.\footnote{
Here we also mention that Fischler and Oliensis
\cite{FO} have derived RGE's for Yukawa coupling parameters
of the minimal SM at two-loop level.}

For several reasons, we considered it instructive to 
perform an independent
derivation of the one-loop RGE's for the scalar and quark
fields and for the Yukawa matrices in the discussed general
2HDM. One reason is that the one-loop results of \cite{Vaughn}
and \cite{MV} do not agree
entirely with those by other authors \cite{JO}-\cite{VD}.
Another reason is that the existing works on the
one- and two-loop RGE's for general (semi)simple gauge
groups $G$ use various conventions for the fermionic
fields, and are usually written in a
language difficult for non-specialists in the method used. 
The third reason is that these works
do not apply the method of finite cutoffs \cite{Lepage}
which we consider especially appealing and
physically intuitive -- although, at two-loops, probably
not the most efficient one.
With our independent cross-check we are confident that
the one-loop results of \cite{JO} are correct.

\section{Numerical examples of evolution}

Here we present a few simple but hopefully typical
examples of the RGE evolution of parameters in the G2HDM.
Some preliminary numerical results were presented
by us in Ref.~\cite{APPB}. For simplicity, we assume:
\begin{itemize} 
\item there is no CP violation -- 
all original four Yukawa matrices 
${\tilde U}^{(j)}$, ${\tilde D}^{(j)}$ are real, and the
VEV phase difference $\eta$ is zero;
\item
the Yukawa parameters of the first
quark generation as well as those of the leptonic sector
are neglected (the quark Yukawa mass matrices
are therefore $2\!\times\!2$).
\end{itemize}

For the boundary conditions to the RGE's, at
the evolution energy $E\!=\!M_Z$,
we first take the CSY ansatz (\ref{FCNCcon1})-(\ref{FCNCcon2}),
with $\xi^{(u)}_{ij}\!=\!1\!=\!\xi^{(d)}_{ij}$ or
$\xi^{(u)}_{ij}\!=\!2\!=\!\xi^{(d)}_{ij}$ 
for all $i,j=1,2$.
We stress that $i\!=\!1$ 
refers now to the second quark family ({\em c,s\/}),
and $i\!=\!2$ to the third family ({\em t,b\/}). 
For the ($2\!\times\!2$) orthogonal CKM mixing matrix $V$ we take
$V_{12}(M_Z)\!=\!0.045\!=\!-V_{21}(M_Z)$. The values of other
parameters at $E\!=\!M_Z$ are chosen to be:\\
$\tan \beta\!=\!1.0$; 
$v\!\equiv\!\sqrt{v_1^2\!+\!v_2^2}\!=\!246.22$ GeV;\\
$\alpha_3\!=\!0.118$, $\alpha_2\!=\!0.332$, 
$\alpha_1\!=\!0.101$; \\
$m_c\!=\!0.77$ GeV, $m_s\!=\!0.11$ GeV, 
$m_b\!=\!3.2$ GeV, and $m_t\!=\!171.5$ GeV.\\ 
The latter quark mass values 
correspond to: $m_c(m_c)\!\approx\!1.3$ GeV,
$m_s(1 {\rm{GeV}})$ $\approx 0.2$ GeV, $m_b(m_b)\!\approx\!4.3$ GeV,
and $m_t^{\rm{phys.}}\!\approx\!174$ GeV
[$m_t(m_t)\!\approx\!166$ GeV]. For ${\alpha}_3(E)$ we
used two-loop evolution formulas, with threshold effect
at $E\!\approx\!m_t^{\rm{phys.}}$ taken into
account; for ${\alpha}_j(E)$ ($j\!=\!1,\!2$) we used one-loop
evolution formulas. 

The described simplified framework resulted in $18$ coupled
RGE's [for $18$ real parameters: $v^2$, $\tan \beta$,
${\tilde U}_{ij}$, ${\tilde D}_{ij}$, ${\tilde G}^{(U)}_{ij}$,
${\tilde G}^{(D)}_{ij}$], with the
mentioned boundary conditions at $E\!=\!M_Z$.
The system of RGE's was solved numerically, using
Runge-Kutta subroutines with adaptive stepsize control
(given in \cite{WHPressetal}). 
The numerical results were cross-checked in several ways,
including the following: FORTRAN programs for the RGE evolution
and for the biunitary transformations were constructed
independently by two of the authors (S.S.H. and G.C.),
and they yielded identical numerical results presented
in this Section.

\begin{figure}[htb]
\mbox{}
\vskip10.5cm\relax\noindent\hskip1.2cm\relax
\includegraphics{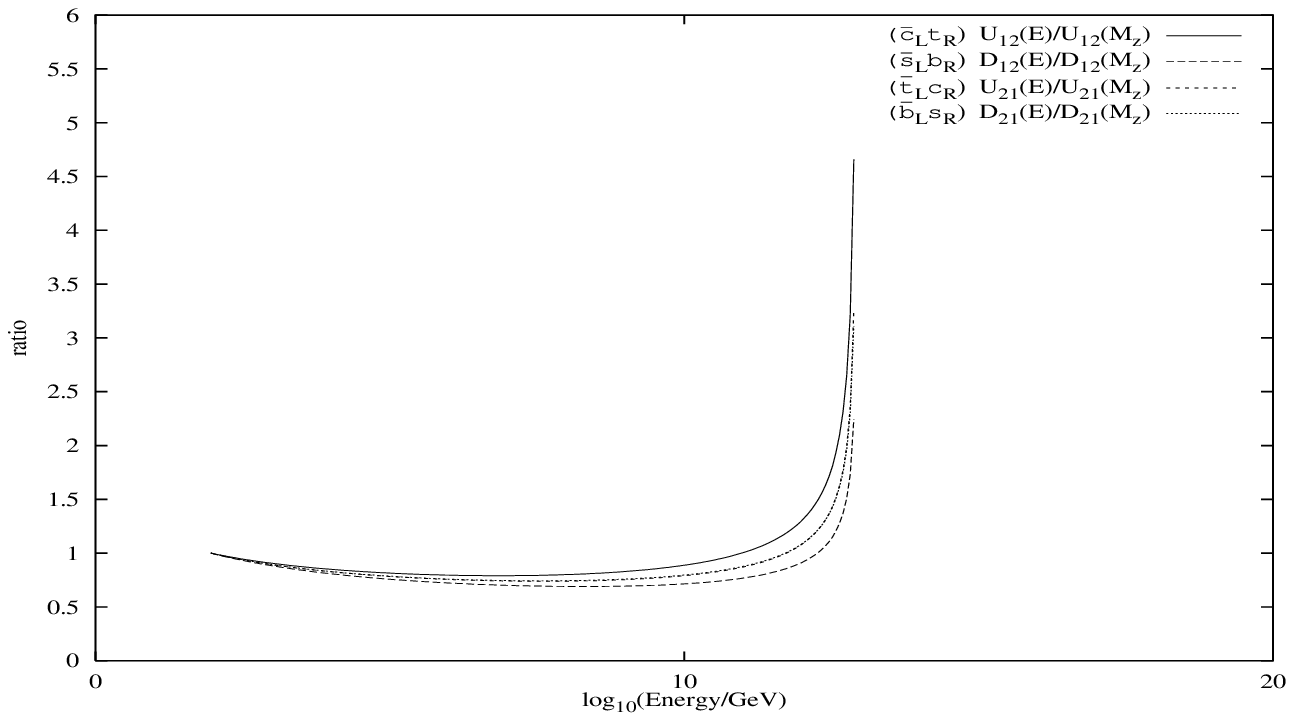} \vskip-3.6cm
\caption{\footnotesize 
FCN Yukawa parameter ratios $U_{ij}(E)/U_{ij}(M_Z)$,
$D_{ij}(E)/D_{ij}(M_Z)$ ($i\!\not=\!j$) in the G2HDM
as the Euclidean energy of probes $E$ increases.
These parameters are in the quark mass basis. 
At $E\!=\!M_Z$, CSY ansatz was taken with
$\xi^{(u)}_{ij}\!=\!\xi^{(d)}_{ij}\!=\!1$
(for all $i,\!j\!=\!1,\!2$).}
\label{xi1}
\end{figure}

\begin{figure}[htb]
\mbox{}
\vskip10.1cm\relax\noindent\hskip1.2cm\relax
\includegraphics{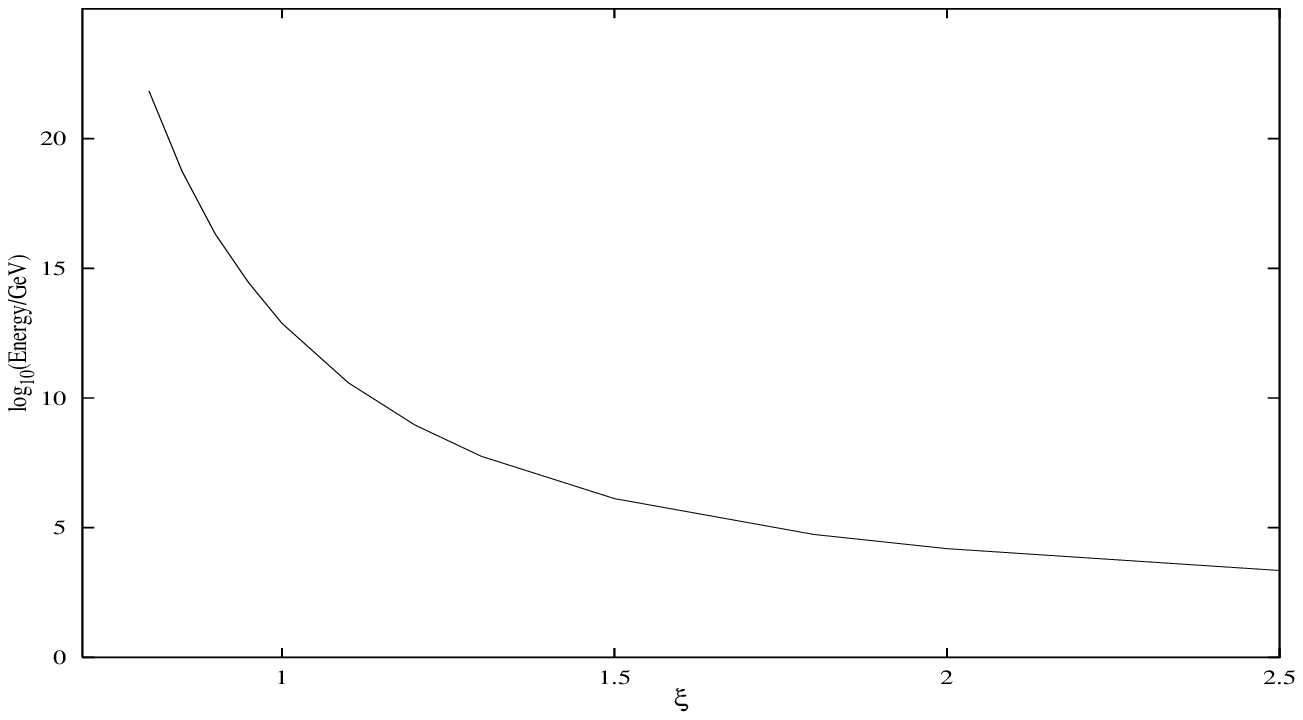} \vskip-3.2cm
\caption{\footnotesize Variation of the Landau pole energy
when the low energy parameters
${\xi}_{ij}^{(u)}\!=\!{\xi}_{ij}^{(d)}\!\equiv\!\xi$ of the
CSY ansatz (\ref{FCNCcon1})-(\ref{FCNCcon2}) are
varied. For $\xi\!=\!2.5$, the onset scale of new physics
is already quite low: $E_{\rm{pole}}\!\approx\!2$ TeV.}
\label{pole}
\end{figure}

The results for the FCN Yukawa parameter ratios
$X_{ij}(E)/X_{ij}(M_Z)$ 
($X\!=\!U,\!D$; $i\!\not=\!j$) are 
given for the case $\xi^{(u)}_{ij}\!=\!1\!=\!\xi^{(d)}_{ij}$ 
in Fig.~\ref{xi1}. 
{}From the
Figure we immediately notice that the FCN coupling 
parameters are remarkably stable as the energy of probes
increases. Even those FCN Yukawa coupling parameters which
involve $t$ quark remain quite stable.
Only very close to the top-quark-dominated
Landau pole 
($E_{\rm{pole}}\!\approx\!0.84 \cdot 10^{13}$ GeV)\footnote{
The value of $E_{\rm{pole}}$
is strongly dependent on the given value of parameter $\xi$, 
as shown later in Fig.~\ref{pole}.}
the coupling parameters start increasing substantially.
For example, in the down-type FCN sector ($b$-$c$)
the corresponding ratio $D_{21}(E)/D_{21}(M_Z)$
acquires its double initial value (\ie{}, value $2$)
at $E\!\approx\!0.7 E_{\rm{pole}}$, which
is very near the (Landau) pole.
About the same holds also for $U_{21}(E)/U_{21}(M_Z)$.
For the ratio $D_{12}(E)/D_{12}(M_Z)$
the corresponding energy is even closer to 
$E_{\rm{pole}}$. For the $t$-quark-dominated
$U_{12}(E)/U_{12}(M_Z)$ it is somewhat lower. 

\begin{figure}[htb]
\mbox{}
\vskip10.1cm\relax\noindent\hskip1.6cm\relax
\includegraphics{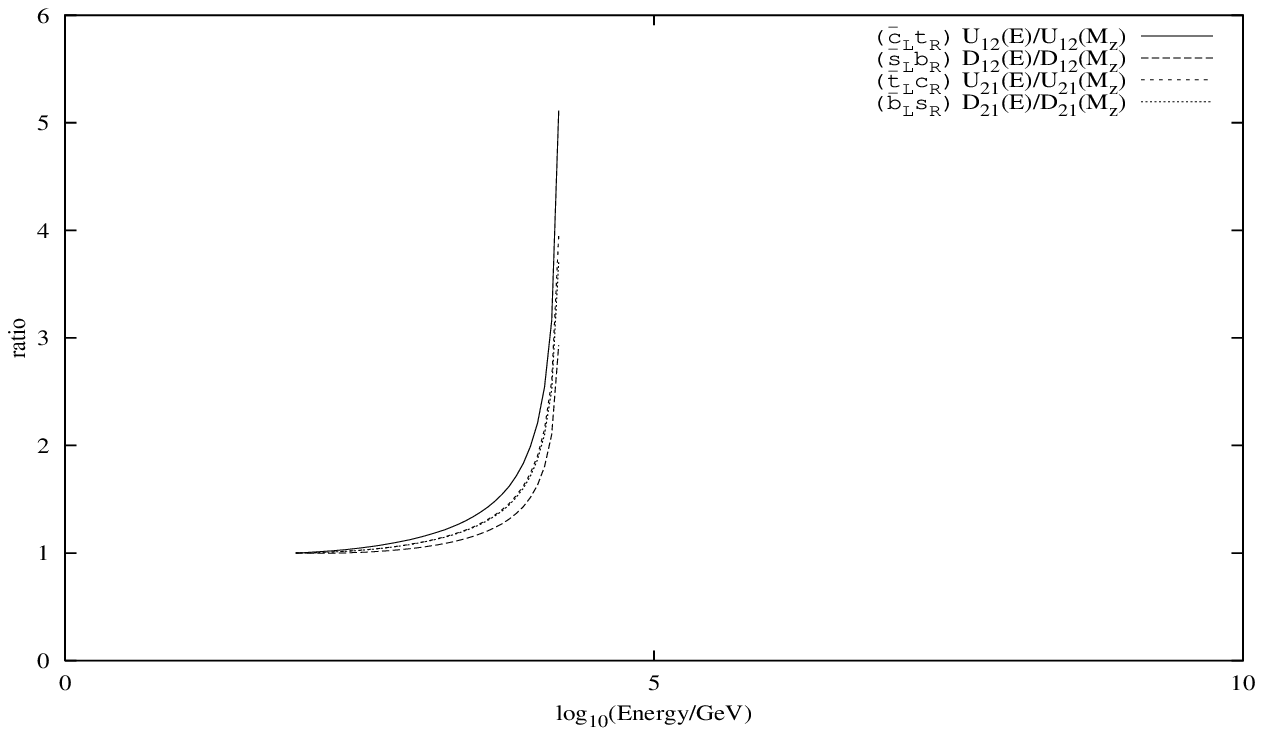} \vskip-3.3cm
\caption{\footnotesize Same as in Fig.~\ref{xi1}, but for the
choice ${\xi}_{ij}^{(u)}\!=\!{\xi}_{ij}^{(d)}\!=2$
(for all $i,\!j\!=\!1,\!2$).}
\label{xi2}
\end{figure}

\begin{figure}[htb]
\mbox{}
\vskip10.1cm\relax\noindent\hskip1.6cm\relax
\includegraphics{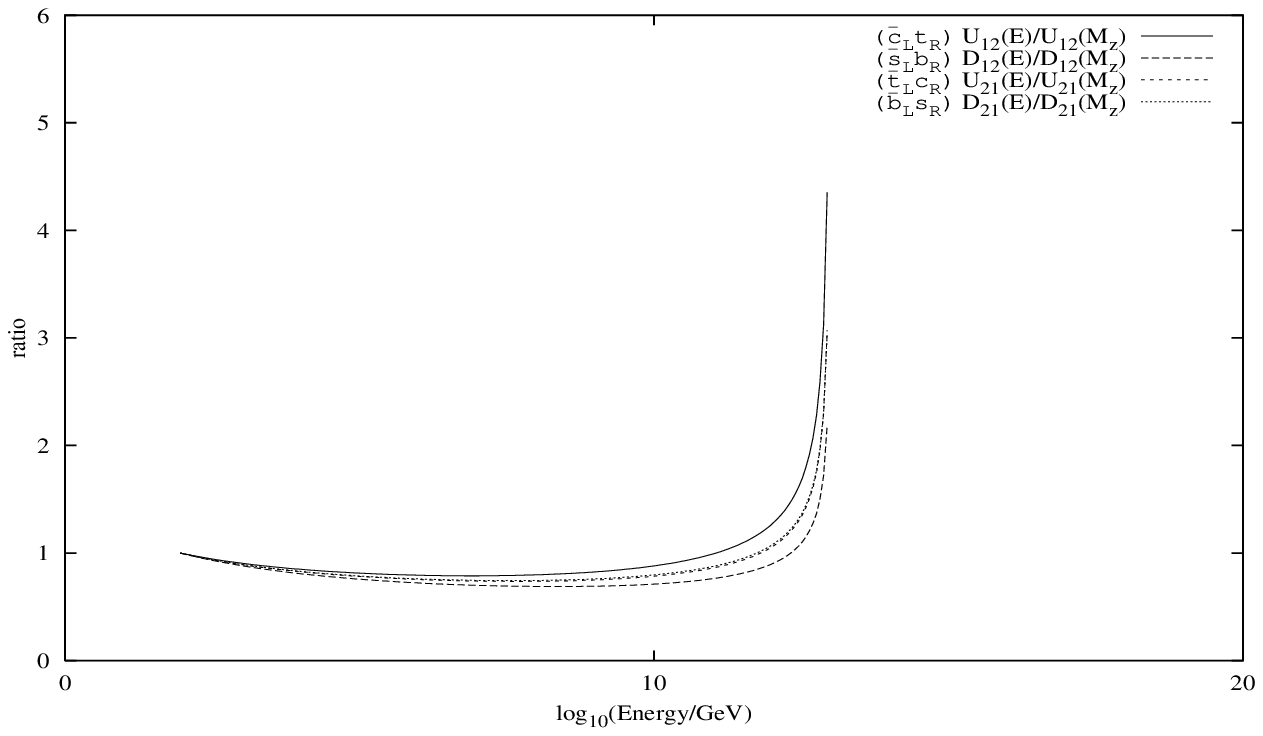} \vskip-3.3cm
\caption{\footnotesize Same as in Fig.~\ref{xi1}, but
for the choice ${\xi}_{12}^{(u)}\!=\!{\xi}_{21}^{(u)}\!=\!0.05163$
(other ${\xi}_{ij}$'s are $1$).}
\label{xisupp}
\end{figure}

In Fig.~\ref{xi2}, evolution of the same FCN ratios
is depicted for the case of the low energy CSY parameters
$\xi^{(u)}_{ij}\!=\!2\!=\!\xi^{(d)}_{ij}$. 
The $t$-quark-dominated Landau pole is now substantially lower
($E_{\rm pole}\!\sim\!10$ TeV), but the behavior of the
FCN ratios remains qualitatively the same. 
Moreover, when some of the CSY parameters
$\xi_{ij}$ are varied, the stability of the FCN ratio
persists, and the Landau pole is influenced almost
entirely by the
$t$-quark-dominated CSY parameter $\xi^{(u)}_{22}$.
We also looked into cases when the CSY ansatz 
is effectively abandoned.
If we suppress the up-type off-diagonal element
at $E\!=\!M_Z$ drastically, for example by taking
$\xi^{(u)}_{12}\!=\!\xi^{(u)}_{21}\!\approx\!0.0516$
[corresponding to $U_{ij}(M_Z)\!=\!D_{ij}(M_Z)$ for $i\!\not=\!j$]
and all other $\xi_{ij}$ parameters equal to $1$, we
obtain results depicted in Fig.~\ref{xisupp}, which are
very close to those of Fig.~\ref{xi1}.

From all these Figures we conclude that 
the FCN Yukawa coupling parameters in the general 2HDM
show remarkable stability when the (Euclidean) energy
of probes increases. This stability persists up to
the energy regions which are, on the logarithmic scale,
quite close to the top-quark-dominated Landau pole.
The general 2HDM appears to possess this stability 
even when the off-diagonal low energy parameters
$|\xi_{ij}|$ (\ref{FCNCcon1})-(\ref{FCNCcon2}) have 
values much smaller than $1$ (while the diagonal parameters
are $\xi_{jj}\!\sim\!1$).

\begin{figure}[htb]
\mbox{}
\vskip10.1cm\relax\noindent\hskip1.2cm\relax
\includegraphics{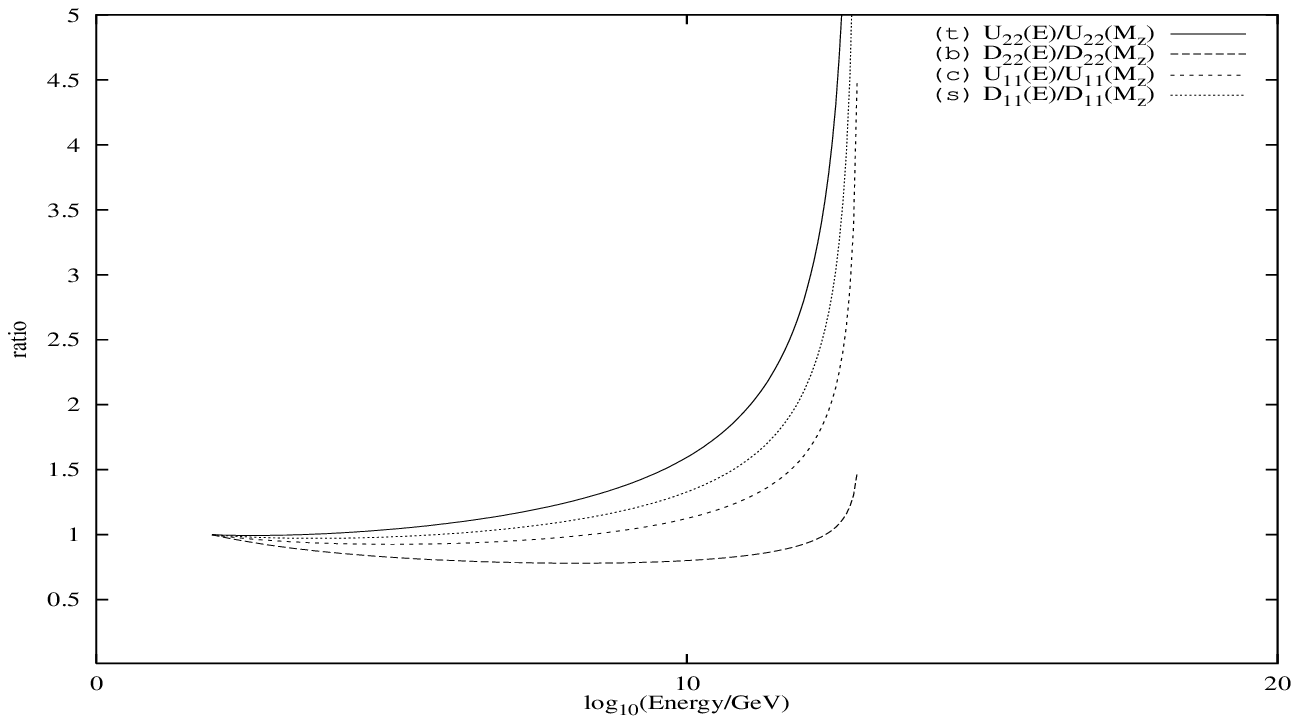} \vskip-3.cm
\caption{\footnotesize Same as in Fig.~\ref{xi1},
but for the neutral current Yukawa coupling parameters 
$U_{jj}$ and $D_{jj}$ ($j\!=\!1,\!2$) which don't change flavor.}
\label{FNNC}
\end{figure}

\begin{figure}[htb]
\mbox{}
\vskip10.1cm\relax\noindent\hskip1.1cm\relax
\includegraphics{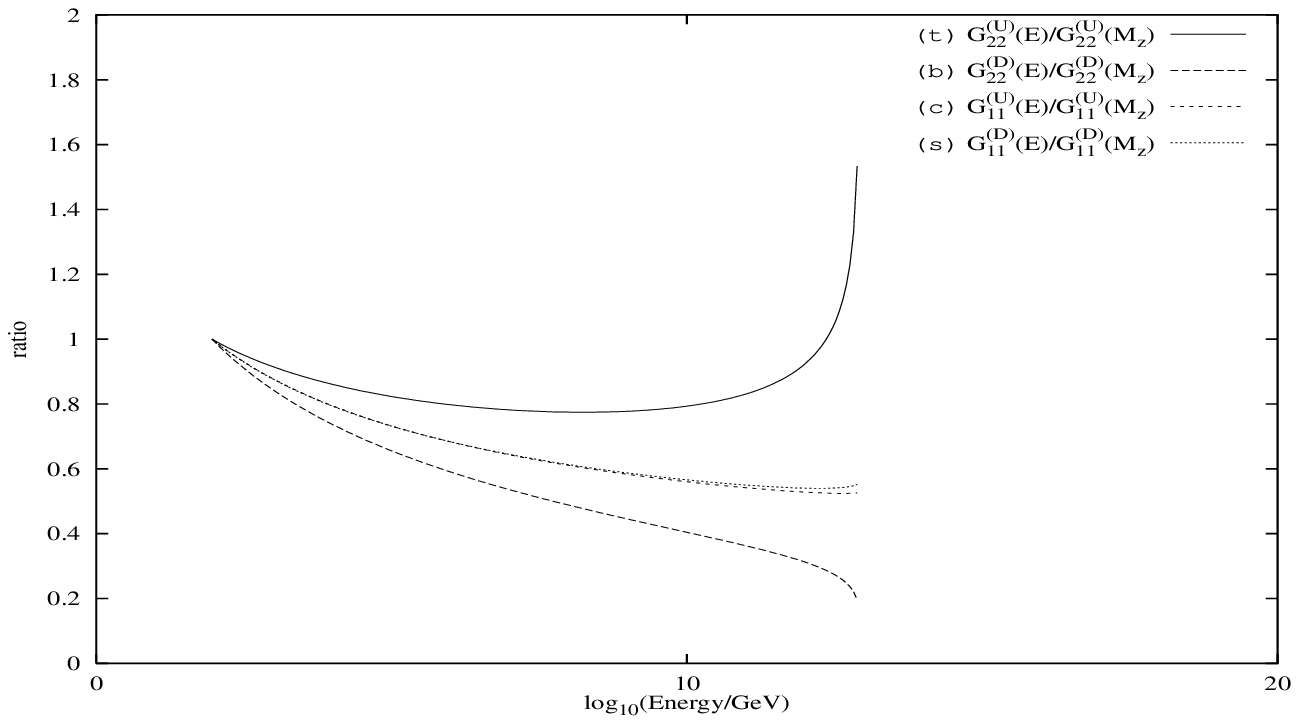} \vskip-3.1cm
\caption{\footnotesize Same as in Fig.~\ref{xi1},
but for the ``mass'' Yukawa parameters $G^{(U)}_{jj}$
and $G^{(D)}_{jj}$ ($j\!=\!1,\!2$) instead.
Since $G^{(U)}(E)$ and $G^{(D)}(E)$ matrices are diagonal
by definition (quark mass basis), these neutral current
Yukawa matrices have zero FCN components
automatically.}
\label{Ymass}
\end{figure}

We can also compare the RGE evolution of the Yukawa
parameters in the G2HDM with those of the MSM,
2HDM(II) and 2HDM(I). 
The authors of \cite{PR}-\cite{HFH} proposed that 
(heavy) quark mass and
Higgs masses in the MSM could be determined by the
infrared fixed points of the RGE's. These questions
were numerically investigated also in different variants
of the 2HDM(II) and 2HDM(I) \cite{HLR}. The authors
of the latter work found out that relatively unambiguous
predictions can be made only if there is a heavy
quark generation and the (heavy) quarks couple to
both Higgs doublets. It is interesting to note that
in the G2HDM, the heaviest ($t$) quark also has 
an infrared fixed point behavior, as suggested from
Figs.~\ref{xi1}-\ref{xisupp}. This is further suggested
from Figs.~\ref{FNNC}-\ref{Ymass} which represent
evolution of the ratios $X(E)/X(M_Z)$
involving the flavor-nonchanging neutral
Yukawa parameters ($X\!=\!U_{jj}$ or $D_{jj}$)
or the ``mass'' Yukawa parameters 
($X\!=\!G^{(U)}_{jj}$ or $G^{(D)}_{jj}$).
By the infrared fixed point behavior we mean
that, for a given approximate 
Landau pole energy $E_{\rm pole}$,
we have very weak sensitivity of $X(M_Z)$ on the
otherwise large value 
$X(E_{\rm pole})\!\stackrel{>}{\sim}\!1$.
The reason for this similarity with the
MSM and 2HDM(II) and 2HDM(I) lies probably
in the conjunction of the facts that the CSY ansatz 
(\ref{FCNCcon1})-(\ref{FCNCcon2}) implies
dominance of the $t$-related Yukawa coupling
parameters and that the QCD contribution
to the evolution of these parameters has the sign
which is in general
opposite to that of the ($t$-related) Yukawa parameter
contributions. Inspecting the work \cite{HLR} further,
we note that it would also be interesting to
investigate the RGE behavior of the quartic Higgs 
coupling parameters in the G2HDM. This would
tell us when these parameters have infrared fixed point
behavior und thus when the physical
Higgs boson mass spectrum can be determined.

In this context, we mention that the idea of RGE
fixed points had been introduced earlier by Chang
\cite{Chang}. He, and subsequently others
\cite{Suzukietal}, investigated connection of
RGE fixed points with asymptotic freedom in
massive gauge theories. Cabibbo {\em et al.\/}
\cite{Cab} were apparently the first to investigate
mass constraints in the minimal SM by imposing
boundary conditions on (perturbative) RGE's
at the unification energy of grand unified theories
[SU(5) or O(10)]. A somewhat related analysis
was performed later by the authors of Ref.~\cite{MV2}
who used SU(5) fixed-point conditions.

It should be stressed that the presented
numerical results are independent of the
chosen value of the VEV ratio, $\tan \beta$, at $E\!=\!M_Z$.
This is connected with our choice of the CSY boundary conditions
(\ref{FCNCcon1})-(\ref{FCNCcon2}) at $E\!=\!M_Z$ for the
Yukawa matrices in the quark mass basis 
(all ${\xi}_{ij}$'s taken real)
and the reality of the chosen CKM matrix at $E\!=\!M_Z$.
These boundary conditions result in real and 
$\beta$-independent Yukawa matrices ${\tilde U}$,
${\tilde D}$, ${\tilde G}^{(U)}$, ${\tilde G}^{(D)}$
in a weak [$\mathrm{SU(2)_L}$] basis\footnote{
We chose at $E\!=\!M_Z$ the following weak basis:
${\tilde U}\!=\!U$, ${\tilde G}^{(U)}\!=\!G^{(U)}$,
${\tilde D}\!=\!V D$, ${\tilde G}^{(D)}\!=\!V G^{(D)}$,
where $V$ is the CKM matrix (at $E\!=\!M_Z$). According to
relations (\ref{Gs})-(\ref{UDs}), the reality of the
Yukawa matrices ${\tilde U}$, ${\tilde D}$, ${\tilde G}^{(U)}$
and ${\tilde G}^{(D)}$ at the low energy $E\!=\!M_Z$ would follow, 
for example, from:
the requirement of no CP violation in the Yukawa sector
(\ie{}, the original Yukawa matrices ${\tilde U}^{(j)}$ and
${\tilde D}^{(j)}$ are all real) {\em together with\/} 
the requirement of no CP violation in the scalar sector 
(\ie{}, the VEV phase difference $\eta=0$) at that low energy.} 
at $E=M_Z$. The RGE's (\ref{RGEU})-(\ref{RGEGD}) then imply that
these matrices remain real and independent of $\beta$
at any evolution energy $E$, and that also their
counterparts $U$, $D$, $G^{(U)}$ and $G^{(D)}$
in the quark mass basis, as well as
the CKM matrix $V$, remain real and independent of $\beta$
at any energy $E$. 
Stated otherwise, if there is
$\beta$-independence and no CP violation 
(neither in the original Yukawa matrices
nor in the scalar sector) at a low energy ($E\!=\!M_Z$),
then these properties persist at all higher energies of 
evolution.\footnote{
CP conservation in the pure scalar sector at a low
energy $E\!=\!M_Z$ (\ie{}, $\eta\!=\!0$) also
persists then at all higher energies of evolution,
since $d \eta/d \ln E\!=\!0$ by the
reality of the Yukawa matrices, according to RGE
(\ref{RGExi}).}

This feature is in stark contrast with the situation in
the 2HDM(II) where the Yukawa matrices strongly depend on
$\beta$ already at low energies -- \eg{}, 
$g_t(M_Z)\!=\!m_t(M_Z) \sqrt{2}/v_u\!=\!m_t(M_Z) \sqrt{2}/[v 
\sin \beta(M_Z)]$. Also the location of the Landau pole
in the 2HDM(II) then crucially depends on $\beta(M_Z)$ --
smaller $\beta(M_Z)$ implies larger $g_t(M_Z)$ and hence
a drastically lower Landau pole.

On the other hand, the G2HDM
treats the up-type and the down-type sectors of quarks 
(the two VEV's $v_1$ and $v_2$) non-discriminatorily.
Therefore, it should be
expected that any reasonable boundary conditions for
Yukawa coupling parameters at low energies should also be independent
of $\beta$ in such frameworks, and this independence
then persists to a large degree
also at higher energies. Also the locations of the
Landau poles (\ie{}, of the approximate scales of 
the onset of new physics)
should then be expected to be largely $\beta$-independent.
In this sense, the G2HDM has more similarity to the 
minimal SM (MSM) than to the 2HDM(II).
The persistence of complete $\beta$-independence of the
Yukawa coupling parameters at high energies
and of the Landau poles, however, can then be ``perturbed'' by
CP violation -- because
RGE's (\ref{RGEU})-(\ref{RGEGD}) are somewhat
$\beta$-dependent when the Yukawa matrices
$\tilde U$, etc., are not real. 

In addition to the connection between (low energy) CP violation
and $\beta$-dependence of
high energy results, there is yet another feature that
distinguishes the G2HDM from the MSM
-- the Landau pole of the G2HDM is in
general much lower than that of the MSM. We can see
that in the following way: let us consider that
only the Yukawa parameters connected with the
top quark degree of freedom are substantial,
\ie{}, $G^{(U)}_{22}\!=\!g_t\!\sim\!1$ and
$U_{22}\!=\!g_t^{\prime}\!\sim\!1$. We have:
$g_t(E)\!=\!m_t(E) \sqrt{2}/v(E)$, as in the MSM, and
$g_t^{\prime}(E)$ is an additional large Yukawa parameter
-- both crucially influence location of the
Landau pole. Inspecting RGE's (\ref{RGEU}) and
(\ref{RGEGU}) for this special approximation
of two variables $g_t$ and $g_t^{\prime}$, we
see that RGE for $g_t$ is similar to that in the
MSM, but with an additional large positive
term on the right: $(3/2) (g_t^{\prime})^2 g_t$.
RGE for $g_t^{\prime}$ has a similar structure as
RGE for $g_t$, but with substantially larger coefficients
at the positive terms on the right. As a result,
$g_t^{\prime}(E)$ is in general larger than
$g_t(E)$. Our specific numerical example
$\xi^{(u)}_{ij}\!=\!\xi^{(d)}_{ij}\!=\!1$
shows that $g_t^{\prime}(E)$ is on average
(average over the whole evolution energy range)
almost twice as large as $g_t(E)$. If we then
simply replace in the mentioned additional
term $(3/2) (g_t^{\prime})^2 g_t$
the parameter $(g_t^{\prime})^2$ by
$3.5 g_t^2$, we obtain from the resulting
``modified'' MSM RGE for $g_t$ a value for the Landau pole
in the region of $10^{12}$-$10^{13}$ GeV, which
is roughly in agreement with the actual value of the Landau
pole of our numerical example 
$E_{\rm{pole}}\!\approx\!0.84 \cdot 10^{13}$ GeV. 
And this value is much lower than $E_{\rm{pole}}$
in the MSM which is above the Planck scale.
As already mentioned, the value of $E_{\rm{pole}}$
is largely influenced by the value of the top-quark-dominated
parameter $\xi^{(u)}_{22}$.
Of course, when we allow the ${\xi}_{ij}$
($\xi^{(u)}_{22}$) parameters of the CSY ansatz 
(\ref{FCNCcon1})-(\ref{FCNCcon2}) at $E\!=\!M_Z$
to deviate from 1, we obtain larger $\log (E_{\rm{pole}})$
for smaller ${\xi}_{ij}$, and smaller
$\log (E_{\rm{pole}})$ for larger
${\xi}_{ij}$. In Fig.~\ref{pole} we depicted this
variation of the Landau pole energy when the
CSY low energy parameters ${\xi}_{ij}$ are varied.

It is also interesting to note that RGE (\ref{RGExi})
for the evolution of the difference of the VEV phases
($\eta$) implies in the G2HDM that $\eta$ can change
when the energy of probes changes. This would
generally occur when we have
CP violation in the Yukawa sector (i.e., complex 
Yukawa matrices). Even when $\eta\!=\!0$
at some low energy, it may become nonzero at some
higher energy due to the CP violation in the
Yukawa sector. This contrasts with the
2HDM(II) or 2HDM(I) where the right side of (\ref{RGExi})
is zero always and thus the CP violation in the Yukawa
sector doesn't influence $\eta$.

\section{Summary and Conclusions}

We performed numerical analysis of the one-loop
RGE's in the general two-Higgs-doublet model (G2HDM).
In the analysis, we neglected the Yukawa coupling
parameters of the light first quark generation,
as well as the contributions of the leptonic sector.
At low energies of probes, we first adopted the CSY ansatz
(\ref{FCNCcon1})-(\ref{FCNCcon2}) which is largely 
motivated by the existing quark mass hierarchies.
We found out that the flavor-changing neutral (FCN)
Yukawa parameters remain remarkably stable when the
energy of probes increases all the way to the
vicinity of the ($t$-quark-dominated) Landau pole.
This conclusion survives even when the CSY ansatz is
effectively abandoned, \ie{}, when the off-diagonal
low energy parameters are additionally
suppressed: $|{\xi}_{ij}^{(u)}|\!\ll\!1$ 
($i\!\not=\!j$). This behavior indicates that
the G2HDM doesn't behave unnaturally with respect
to the RGE evolution of the vertices of the Higgs-exchanged
flavor-changing neutral currents. 
Since the G2HDM, in contrast to the 2HDM(II) and 2HDM(I),
has no explicit and exact 
discrete [or $\mathrm{U(1)}$] family symmetries
which would ensure persistence of the FCN Yukawa
suppression at increasing energies of probes,
the behavior of FCN Yukawa parameters found numerically
in the present paper may be somewhat surprising.
The general suspicion about the G2HDM in the past
had centered on the fact that absence of the mentioned
family symmetries in the Lagrangian density would
in general not keep FCN Yukawa parameters suppressed
under the RGE evolution and would thus render the
model unnatural and fraught with fine-tuning of
``bare'' FCN Yukawa parameters. Stated otherwise,
the RGE's of the G2HDM would in general allow a ``pull-up''
effect by the diagonal Yukawa parameters
on the much smaller off-diagonal (FCN) ones.
This would increase the values of the latter by a large factor
or even by orders of magnitude when the energy of probes
increases by one or several orders of magnitude.
Our numerical analysis shows that this doesn't happen,
at least as long as the low energy parameters of the
CSY ansatz (\ref{FCNCcon1})-(\ref{FCNCcon2})
safisfy $|{\xi}_{ij}|\!\stackrel{<}{\sim}\!1$
for $i\!\not=\!j$ and ${\xi}_{jj}\!\sim\!1$.
Perhaps it is the latter condition (for the diagonal
$\xi$ parameters) which causes the mentioned
persistence of the FCN Yukawa suppression.
The latter condition, together with the
known form of the CKM matrix, effectively
represents an approximate symmetry in which
only the third quark generation has substantially nonzero
Yukawa parameters (and almost no CKM mixing). This
can be called also approximate flavor democracy.

Further, 
the high energy Yukawa coupling parameters
in the model have in general little dependence 
on the VEV ratio $\tan \beta$ as long as
the CP violation is weak.
Moreover, we found out that the G2HDM has an interesting
behavior of the Landau pole energies. 
They can become quite low ($\sim\!1$ TeV) already
at not very high $\xi_{ij}$ parameters 
(${\xi}_{22}^{(u)}\!\leq\!3$), as shown in Fig.~\ref{pole}. 
These energies, signaling the breakdown of the
perturbative approach in the G2HDM, can be interpreted 
as possible scales of the onset of a new physics and/or
a strong coupling regime.

\vspace{1.cm}

{\noindent{\bf Abbreviations used in the article:}}\\

{\noindent
AHR -- Antaramian, Hall and Ra\v sin;}\\
{\noindent
CKM -- Cabibbo, Kobayashi and Maskawa;}\\
{\noindent
CSY -- Cheng, Sher and Yuan;}\\
{\noindent
FCN -- flavor-changing neutral;}\\
{\noindent
FCNC -- flavor-changing neutral current;}\\
{\noindent
G2HDM -- general two-Higgs-doublet (Standard) model;}\\
{\noindent
MSM -- minimal Standard Model;}\\
{\noindent
RGE -- renormalization group equation;}\\
{\noindent
VEV -- vacuum expectation value;}\\
{\noindent
1PI -- one-particle-irreducible;}\\
{\noindent
1PR -- one-particle-reducible;}\\
{\noindent
2HDM -- two-Higgs-doublet (Standard) Model.}\\

\vspace{1.cm}

{\noindent{\bf Acknowledgments:}}\\

\vspace{0.3cm}

\noindent
G.C.~wishes to acknowledge the hospitality and 
financial support of Bielefeld University and DESY Hamburg, 
where part of this work was done.
C.S.K.~wishes to acknowledge the financial support
of the Korean Research Foundation.
The work of S.S.H.~was supported in part by the
Non-Directed-Research-Fund KRF, in part by the CTP,
Seoul National University, in part by the BSRI Program,
Ministry of Education, Project No.~BSRI-98-2425, and in 
part by the KOSEF-DFG large collaboration project, 
Project No.~96-0702-01-01-2.
\newpage

\begin{appendix}

\section[]{One-loop RGE's in the general 2HDM}
\setcounter{equation}{0}
We outline here a derivation of the one-loop RGE's for the
scalar and quark fields and for the Yukawa coupling matrices
${\tilde D}^{(k)}$ and ${\tilde U}^{(k)}$ ($k\!=\!1,\!2$)
in the general two-Higgs doublet model (G2HDM) whose
Lagrangian density in the Yukawa sector is represented
by (\ref{2HD30}). The derivation follows the finite cutoff
interpretation of RGE's as presented, for example, in 
Ref.~\cite{Lepage}. We consider it useful to present the
derivation because the approach used here is
physically very intuitive, and because
the existing relevant literature on
RGE's in general (semi)simple Lie gauge groups is written
in a rather cryptic manner and not all the works agree
completely with each other. In Sec.~3 we also compared
the results obtained here with those implied
by the existing literature.

\subsection[]{{\bf One-loop RGE's for the scalar fields}}

\begin{figure}[htb]
\mbox{}
\vskip4.cm\relax\noindent\hskip3.1cm\relax
\includegraphics{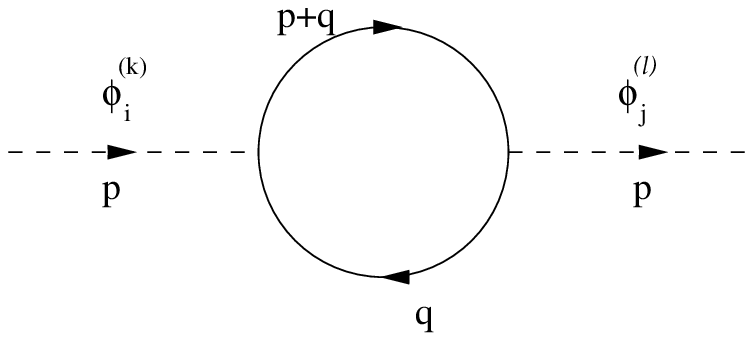} \vskip-0.2cm
\caption{\footnotesize The diagram leading to the
two-point Green function 
$-{\mathrm{i}} \Sigma_{ij}^{(k,\ell)}(p^2; E^2)$.
Full lines represent quark propagators.}
\label{sc2pG}
\end{figure}

To obtain evolution of the scalar fields
${\phi}_i^{(k)}(E)$ with ``cutoff'' energy $E$,
the truncated (one-loop) two-point Green
functions $-{\mathrm{i}} \Sigma_{ij}^{(k,\ell)}(p^2; E^2)$, 
represented diagrammatically in Fig.~\ref{sc2pG},
have to be calculated first.
More specifically, it suffices to calculate only
their cutoff-dependent parts $\propto\!p^2 \ln E^2$ 
which are responsible for effective\footnote{
For clearer notation, we denote in this Section the evolving
(UV cutoff) energy $E$ at the fields not as a superscript, but
rather as an argument.}
kinetic-energy-type
terms $\sim\!{\partial}_{\nu}{\phi}_i^{(k)}(E) 
{\partial}^{\nu} {\phi}_j^{(\ell)}(E)$. In the course of the
calculations, all the masses $m \sim E_{\mathrm{ew}}$
of the relevant particles in the diagrams are ignored. 
This would correspond to
the picture with a finite but large ultraviolet
energy cutoff $E\!\gg\!E_{\rm{ew}}$. Therefore,
calculations need not be performed 
in the mass basis of the relevant particles.
These particles are regarded as effectively massless in the
approximation, the transformations between the original
bases of the relevant fields and their mass bases
are unitary, and therefore the (mass-independent parts of the)
calculated Green functions are the same in both bases.

\begin{figure}[htb]
\mbox{}
\vskip3.cm\relax\noindent\hskip1.3cm\relax
\includegraphics{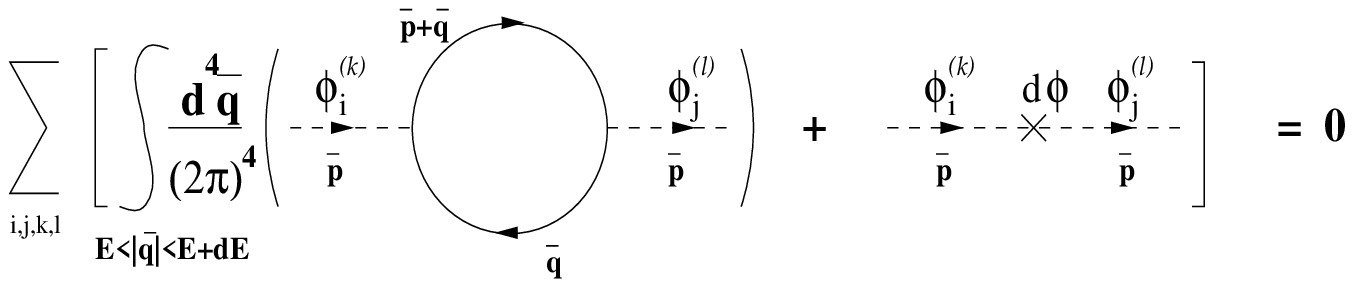} \vskip0.cm
\caption{\footnotesize Diagrammatic illustration of the
RGE relation (\ref{kinLep}) leading to the evolution of the
scalar fields. ${\phi}_i^{(k)}$ stands for ${\phi}_i^{(k)}(E)$,
and $d {\phi}$ stands for ${\phi}(E\!+\!dE)\!-\!{\phi}(E)$
($\phi$ is a generic notation for ${\phi}_j^{(\ell)}$'s). 
The cross represents the contribution of the change of the 
kinetic energy terms originating
from the changes $d {\phi}$ of scalar fields.}
\label{evolsc}
\end{figure}

Calculation of the Green functions
$- {\mathrm{i}} {\Sigma}_{i,j}^{(k,\ell)}(p^2; E^2)$
is then straightforward. The relevant massless integrals
over internal quark-loop momenta $q$ can be carried out
in the Euclidean metric [${\bar q} =
(-{\mathrm{i}} q^0, -q^j)$, 
${\bar p} = (- {\mathrm{i}} p^0, -p^j)$], 
where the upper bound in the loop integral
is: ${\bar q}^2\!\leq\!E^2$. After rotating back into
Minkowski metric (${\bar p}^2\!\mapsto\!-p^2$), we obtain:
\begin{enumerate}
\item 
Green functions whose external legs ${\phi}_i^{(k)}$ and 
${\phi}_j^{(\ell)}$ have the same scalar indices ($i\!=\!j$):
\begin{eqnarray}
\lefteqn{
- {\mathrm{i}} {\Sigma}_{j,j}^{(k,\ell)}(p^2;E^2)  =  
{\mathrm{i}} \frac{ N_{\rm c} }{ 32 {\pi}^2 } p^2 
\ln \left( \frac{E^2}{m^2} \right) \times }
\nonumber\\
&& 
\,\,\,\,\,\,\,
\times {\mathrm{Tr}} 
\left[ {\tilde U}^{(k)} {\tilde U}^{(\ell) \dagger} +
{\tilde U}^{(\ell)} {\tilde U}^{(k) \dagger} +
{\tilde D}^{(k)} {\tilde D}^{(\ell) \dagger} +
{\tilde D}^{(\ell)} {\tilde D}^{(k) \dagger} \right] (E) \ ,
\label{Sigkk}
\end{eqnarray}
where $j\!=\!1,\!2,\!3,\!4$ (no running over $j$); 
$k, \ell\!=\!1,\!2$; 
$m$ is an arbitrary but fixed mass 
($m\!\sim\!E_{\rm{ew}}$).
\item
Green functions whose external legs ${\phi}_j^{(k)}$ and 
${\phi}_{j^{\prime}}^{(\ell)}$ have complementary 
scalar indices
$(j j^{\prime})\!=\!(12)$, $(21)$, $(34)$, $(43)$:
\begin{eqnarray}
\lefteqn{
- {\mathrm{i}} {\Sigma}_{ (j,j^{\prime}) }^{k,\ell}(p^2; E^2) =
(-1)^j \frac{ N_{\rm c} }{ 32 {\pi}^2 } 
p^2 \ln \left( \frac{E^2}{m^2} \right) 
\times }
\nonumber\\
&& 
\,\,\,\,\,\,\,
\times {\mathrm{Tr}} 
\left[ {\tilde U}^{(k)} {\tilde U}^{(\ell)\dagger} - 
{\tilde U}^{(\ell)} {\tilde U}^{(k)\dagger} - 
{\tilde D}^{(k)} {\tilde D}^{(\ell)\dagger} + 
{\tilde D}^{(\ell)} {\tilde D}^{(k)\dagger} \right] (E) \ .
\label{Sigkl1}
\end{eqnarray}
\item
Green functions $-{\mathrm{i}} {\Sigma}_{i,j}^{(k,\ell)}$
with other indices are zero.
\end{enumerate}
All these Green functions can be induced 
alternatively at the tree level
by kinetic energy terms. For example, in theory with
the UV cutoff $E$, 
the kinetic energy term ${\partial}_{\nu} {\phi}_i^{(k)}(E)
{\partial}^{\nu} {\phi}_j^{(\ell)}(E)$ induces (at the tree level)
the two-point Green
function value $-{\mathrm{i}} 
{\Sigma}_{i,j}^{(k,\ell)}(p^2; E^2)\!=\!{\mathrm{i}} p^2$ 
if ${\phi}_i^{(k)}(E)\!\not=\!{\phi}_j^{(\ell)}(E)$, 
and the value $2 {\mathrm{i}} p^2$ if 
${\phi}_i^{(k)}(E)\!\equiv\!{\phi}_j^{(\ell)}(E)$. 
Now, following the finite-cutoff interpretation of RGE's
as described, for example, in Ref.~\cite{Lepage},
we compare the kinetic energy terms in the theory with
the UV cutoff $E$ and in the equivalent theory with the slightly 
different cutoff $(E\!+\!dE)$. The two-point Green
functions in these two equivalent theories must be identical.
Imposition of this requirement in the tree plus one-loop
approximation then leads to the following relation:
\begin{eqnarray}
\lefteqn{
\frac{1}{2} \sum_{j=1}^4 \sum_{k=1}^2 
{\partial}_{\nu} {\phi}_j^{(k)}(E)
{\partial}^{\nu} {\phi}_j^{(k)}(E)
= \frac{1}{2} \sum_{j=1}^4 \sum_{k=1}^2 
{\partial}_{\nu} {\phi}_j^{(k)}(E\!+\!dE)
{\partial}^{\nu} {\phi}_j^{(k)}(E\!+\!dE) +}
\nonumber\\
&& 
\,\,\,\,\,\,\,\,\,\,\,\,\,\,\,\,\,
\,\,\,\,\,\,\,\,\,\,\,\,\,\,\,\,\,
 + \frac{ N_{\rm c} }{ 32 {\pi}^2 } ( d \ln E^2 ) {\Bigg \{}
\sum_{j=1}^4 \sum_{k,\ell=1}^2 A_{k \ell}(E) 
{\partial}_{\nu} {\phi}_j^{(k)}(E)
{\partial}^{\nu} {\phi}_j^{(\ell)}(E) +
\nonumber\\
&& 
\,\,\,\,\,\,\,\,\,\,\,\,\,\,\,\,\,
\,\,\,\,\,\,\,\,\,\,\,\,\,\,\,\,\,
+  \sum_{ (j,j^{\prime}) } \sum_{k,\ell=1}^2 
(-1)^{ j^{\prime} + {\ell} } B_{k \ell}(E)
{\partial}_{\nu} {\phi}_j^{(k)}(E)
{\partial}^{\nu} {\phi}_{ j^{\prime} }^{(\ell)}(E)
{\Bigg \}} \ ,
\label{kinLep}
\end{eqnarray}
where: summation in the last sum runs over already mentioned
complementary indices
$(j j^{\prime})\!=\!(12)$, $(21)$, $(34)$, $(43)$;
$d (\ln E^2)\!\equiv\!\ln(E\!+\!d E)^2\!-\!\ln E^2\!=\!2 d E/E$; 
and elements of the real symmetric matrices
$A(E)$ and $B(E)$ are related to the
Green function expressions
(\ref{Sigkk}) and (\ref{Sigkl1}), respectively:
\begin{eqnarray}
\!\!\!\!\!\!\!\!\!\!
A_{k \ell}(E) &=& \frac{1}{2} {\mathrm{Tr}} \left[
{\tilde U}^{(k)} {\tilde U}^{(\ell)\dagger}\!+\!
{\tilde U}^{(\ell)} {\tilde U}^{(k)\dagger}\!+\!
{\tilde D}^{(k)} {\tilde D}^{(\ell)\dagger}\!+\!
{\tilde D}^{(\ell)} {\tilde D}^{(k)\dagger}  \right](E) \ ,
\label{As}
\\
\!\!\!\!\!\!\!\!\!\!
B_{k \ell}(E) 
& = &(-1)^{\ell} \frac{\rm i}{2} {\mathrm{Tr}} \left[
{\tilde U}^{(k)} {\tilde U}^{(\ell)\dagger}\!-\!
{\tilde U}^{(\ell)} {\tilde U}^{(k)\dagger}\!-\!
{\tilde D}^{(k)} {\tilde D}^{(\ell)\dagger}\!+\!
{\tilde D}^{(\ell)} {\tilde D}^{(k)\dagger}  \right](E) \ .
\label{Bs}
\end{eqnarray}
Equation (\ref{kinLep}) is described in the following way:
the double sum on the left
and the first double sum on the right represent 
the kinetic energy terms of the
scalars in the formulation with UV cutoff $E$ 
and $(E\!+\!d E)$, respectively. 
The one-loop contributions of Fig.~\ref{sc2pG}
with the loop momentum $|{\bar q}|$ in the 
Euclidean energy interval
$E\!\leq\!|{\bar q}|\!\leq\!{\Lambda}$ 
are already contained in the
kinetic energy terms of the left effectively at the tree level
(${\Lambda}$ is a large cutoff where the theory is presumed to
break down). On the other hand,
the kinetic energy terms of the $(E\!+\!d E)$
cutoff formulation [the first double 
sum on the right of (\ref{kinLep})] 
effectively contain, at the tree level, the one-loop 
effects of Fig.~\ref{sc2pG}
for the slightly smaller energy interval: 
$(E\!+\!d E)\!\leq\!|{\bar q}|\!\leq\!{\Lambda}$. 
Therefore, the Green function contributions\footnote{
More precisely: the corresponding effective kinetic energy terms.}
$- {\mathrm{i}} d {\Sigma}_{i,j}^{(k,\ell)}(p^2;E^2)$
of Fig.~\ref{sc2pG} from the loop-momentum interval 
$E\!\leq\!|{\bar q}|\!\leq\!(E\!+\!dE)$ had to be included
on the right of (\ref{kinLep}) -- these are the terms in the
last two double sums there. This is 
illustrated in Fig.~\ref{evolsc}.

In order to find RGE's for the scalar fields ${\phi}_j^{(k)}(E)$,
we make the following ansatz for solution of Eq.~(\ref{kinLep}):
\begin{eqnarray}
{\vec \phi}_j(E\!+\!d E) 
&= &{\vec \phi}_j(E) +
{d \alpha}^{(j)}(E) {\vec \phi}_j(E) +
 {d \beta}^{(j)}(E) {\vec \phi}_{ j^{\prime} }(E) \ ,
\label{ansatz}
\end{eqnarray}
where ${\vec \phi}_j$ is two-component column made up of
${\phi}_j^{(1)}$ and ${\phi}_j^{(2)}$;
$d {\alpha}^{(j)}(E)$ and $d {\beta}^{(j)}(E)$ 
are infinitesimally small $2\!\times\!2$ matrices, and
$d {\beta}^{(j)}$ has zero diagonal elements;
and scalar indices $(jj^{\prime})$ are again complementary.
Inserting ansatz (\ref{ansatz}) into RGE relation (\ref{kinLep}),
we obtain relations
\begin{eqnarray}
{d \alpha}_{k \ell}^{(j)}(E) + {d \alpha}_{\ell k}^{(j)}(E) 
&=& - N_{\rm c} \left( d \ln E^2 \right) 
A_{k \ell}(E)/(16 \pi^2) \ ,
\label{delalph}
\\
{d \beta}_{k \ell}^{(j)}(E) + 
{d \beta}_{\ell k}^{ (j^{\prime}) }(E)
& =&
(-1)^{j+\ell} N_{\rm c} \left( d \ln E^2 \right) 
B_{k \ell}(E)/(16 \pi^2) \ .
\label{delbet}
\end{eqnarray}
In principle, these relations alone do not define the
elements ${d \alpha}_{k \ell}^{(j)}(E)$ and
${d \beta}_{k \ell}^{(j)}(E)$. However, 
RGE evolution of the isodoublet fields
${\Phi}^{(1)}(E)$ and ${\Phi}^{(2)}(E)$ should be invariant
under the exchange of Higgs generation indices 
$1\!\leftrightarrow\!2$,
because these two Higgs doublets appear in the original
Lagrangian density (\ref{2HD30}) in a completely 
$1\!\leftrightarrow\!2$ symmetric manner. 
We will see in retrospect
that this exhange symmetry is respected once we impose the
conditions 
\begin{equation}
{d \alpha}_{k \ell}^{(j)}(E) = {d \alpha}_{\ell k}^{(j)}(E) \ ,
\quad
{d \beta}_{k \ell}^{(j)}(E) = 
{d \beta}_{\ell k}^{ (j^{\prime}) }(E) \ .
\label{symm}
\end{equation}
Relations (\ref{delalph})-(\ref{symm}) 
lead to specific expressions for the evolution
coefficients ${d \alpha}_{k \ell}^{(j)}(E)$
and ${d \beta}_{k \ell}^{(j)}(E)$. When 
inserting these coefficients
back into ansatz (\ref{ansatz}),
we obtain the one-loop RGE's for the evolution of the scalar
fields:
\begin{eqnarray}
\lefteqn{
\frac{16 \pi^2}{N_{\mathrm{c}}}
\frac{d}{d \ln E} {\phi}_j^{(1)}(E) =
 - {\mathrm{Tr}} \left[ 
{\tilde U}^{(1)} {\tilde U}^{(1)\dagger} +
{\tilde D}^{(1)} {\tilde D}^{(1)\dagger} 
\right] {\phi}_j^{(1)} }
\nonumber\\
&&
- \frac{1}{2} {\mathrm{Tr}} \left[ 
{\tilde U}^{(1)} {\tilde U}^{(2)\dagger} +
{\tilde U}^{(2)} {\tilde U}^{(1)\dagger} +
{\tilde D}^{(1)} {\tilde D}^{(2)\dagger} +
{\tilde D}^{(2)} {\tilde D}^{(1)\dagger}\right] {\phi}_j^{(2)} 
\nonumber\\
&&
+ {\mathrm{i}} (-1)^j \frac{1}{2} {\mathrm{Tr}} \left[ 
{\tilde U}^{(1)} {\tilde U}^{(2)\dagger} -
{\tilde U}^{(2)} {\tilde U}^{(1)\dagger} -
{\tilde D}^{(1)} {\tilde D}^{(2)\dagger} +
{\tilde D}^{(2)} {\tilde D}^{(1)\dagger}\right] 
{\phi}_{j^{\prime}}^{(2)} ,
\label{RGEphi1Y}
\end{eqnarray}
\begin{eqnarray}
\lefteqn{
\frac{16 \pi^2}{N_{\mathrm{c}}}
\frac{d}{d \ln E} {\phi}_j^{(2)}(E) =
 - {\mathrm{Tr}} \left[ 
{\tilde U}^{(2)} {\tilde U}^{(2)\dagger} +
{\tilde D}^{(2)} {\tilde D}^{(2)\dagger} \right] {\phi}_j^{(2)} }
\nonumber\\
&&
- \frac{1}{2} {\mathrm{Tr}} \left[ 
{\tilde U}^{(1)} {\tilde U}^{(2)\dagger} +
{\tilde U}^{(2)} {\tilde U}^{(1)\dagger} +
{\tilde D}^{(1)} {\tilde D}^{(2)\dagger} +
{\tilde D}^{(2)} {\tilde D}^{(1)\dagger}\right] {\phi}_j^{(1)} 
\nonumber\\
&&
+ {\mathrm{i}} (-1)^{j+1} \frac{1}{2} {\mathrm{Tr}} \left[ 
{\tilde U}^{(1)} {\tilde U}^{(2)\dagger} -
{\tilde U}^{(2)} {\tilde U}^{(1)\dagger} -
{\tilde D}^{(1)} {\tilde D}^{(2)\dagger} +
{\tilde D}^{(2)} {\tilde D}^{(1)\dagger}\right] 
{\phi}_{j^{\prime}}^{(1)} ,
\label{RGEphi2Y}
\end{eqnarray}
where again $j^{\prime}$ is the scalar 
index complementary to index
$j$: $(j j^{\prime})\!=\!(12)$, $(21)$, $(34)$, $(43)$.
These RGE's lead to RGE's for scalar isodoublets
${\Phi}^{(k)}$:
\begin{equation}
\frac{16 \pi^2}{N_{\mathrm{c}}}
\frac{d}{d \ln E} {\Phi}^{(k)}(E) =
 - \sum_{\ell = 1}^2 {\mathrm{Tr}} \left[ 
{\tilde U}^{(k)} {\tilde U}^{(\ell)\dagger} +
{\tilde D}^{(\ell)} {\tilde D}^{(k)\dagger} 
\right] {\Phi}^{(\ell)}
\ .
\label{RGEPhiY}
\end{equation}
We really see that this set of 
one-loop RGE's is invariant under
the exchange $1\!\leftrightarrow\!2$, 
as required by the form of the
Yukawa Lagrangian density (\ref{2HD30}).

In addition to quark loops, there are also
loops of the electroweak gauge bosons
contributing to one-loop two-point Green functions of the
scalars. However, since these gauge bosons
couple to the Higgs isodoublets identically as in 
the minimal Standard Model (MSM), 
their contributions\footnote{
They are gauge dependent.}
to the right of RGE's (\ref{RGEphi1Y})-(\ref{RGEphi2Y})
and (\ref{RGEPhiY})
are the same as in the MSM.\footnote{
For these contributions
of EW gauge bosons in the MSM, see for example Arason {\em et al.}
\cite{Arasonetal}, App.~A. However, note that they use
for the ${\mathrm{U(1)_Y}}$ gauge coupling $g_1$ a different, 
GUT-motivated, convention:
$(g_1^2)_{\mathrm{Arason et al.}} = 
(5/3) (g_1^2)_{\mathrm{here}}$. } 
Hence, the full one-loop
RGE's for the the scalar isodoublets in the G2HDM, in the
Landau gauge, are
\begin{eqnarray}
16 \pi^2 \frac{d}{d \ln E} {\Phi}^{(k)}(E) &=&
 - N_{\mathrm{c}} \sum_{\ell = 1}^2 {\mathrm{Tr}} \left[ 
{\tilde U}^{(k)} {\tilde U}^{(\ell)\dagger} +
{\tilde D}^{(\ell)} {\tilde D}^{(k)\dagger} \right] {\Phi}^{(\ell)}
\nonumber\\
&&+ \left[ \frac{3}{4} g_1^2(E) + \frac{9}{4} g_2^2(E) \right]
{\Phi}^{(k)}(E) \ ,
\label{RGEPhi}
\end{eqnarray}
and completely analogous gauge boson contributions have to be added
also on the right of (\ref{RGEphi1Y})-(\ref{RGEphi2Y}).
These RGE's are simultaneously also RGE's for the corresponding
VEV's (\ref{2HDnot3})
\begin{eqnarray}
16 \pi^2 \frac{d}{d \ln E} \left( 
{\mathrm{e}}^{{\mathrm{i}} \eta_k} v_k \right) & = &
 - N_{\mathrm{c}} \sum_{\ell = 1}^2 {\mathrm{Tr}} \left[ 
{\tilde U}^{(k)} {\tilde U}^{(\ell)\dagger} +
{\tilde D}^{(\ell)} {\tilde D}^{(k)\dagger} \right] 
\left( {\mathrm{e}}^{{\mathrm{i}} \eta_{\ell}} v_{\ell} \right)
\nonumber\\
&&+ \left[ \frac{3}{4} g_1^2(E) + \frac{9}{4} g_2^2(E) \right]
\left( {\mathrm{e}}^{{\mathrm{i}} \eta_k} v_k \right) \ .
\label{RGEVEVs}
\end{eqnarray}

In this paper we don't discuss the question of quadratic
cutoff terms ${\Lambda}^2$ which appear in the radiative
corrections to VEV's in any SM framework. In the MSM,
their consideration -- under the assumption of the top quark 
dominance of the radiative corrections in the scalar sector -- 
leads to severe upper bounds on the ultraviolet cutoff ${\Lambda}$
for a substantial subset of values of the
bare doublet mass and of the bare scalar self-interaction 
parameters $M^2({\Lambda})$ and $\lambda({\Lambda})$ --
cf.~Ref.~\cite{Fateloetal}.

In order to derive one-loop RGE's for the 
Yukawa matrices ${\tilde U}^{(k)}$
and ${\tilde D}^ {(k)}$, the results
(\ref{RGEphi1Y})-(\ref{RGEPhi}) are needed.
In addition, 
RGE's for evolution of the quark fields 
${\tilde u}^ {(j)}_{L,R}$ and ${\tilde d}^ {(j)}_{L,R}$
are also needed.

\subsection[]{{\bf One-loop RGE's for the quark fields}}
 
\begin{figure}[htb]
\mbox{}
\vskip3.cm\relax\noindent\hskip1.9cm\relax
\includegraphics{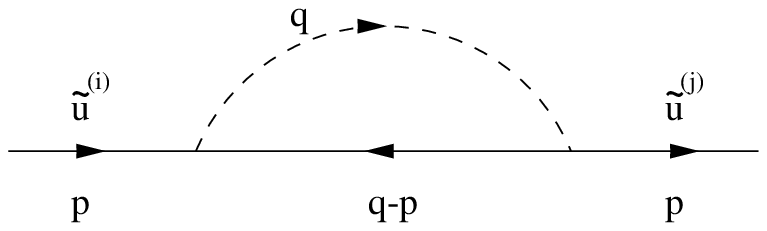} \vskip0.1cm
\caption{\footnotesize The diagram leading to the
two-point Green function 
$-{\mathrm{i}} 
\Sigma( p; E; {\tilde u}^{(i)}, {\tilde u}^{(j)} )$.
Dashed and full lines represent scalar and quark propagators, 
respectively.}
\label{q2pG}
\end{figure}

\begin{figure}[htb]
\mbox{}
\vskip3.cm\relax\noindent\hskip1.5cm\relax
\includegraphics{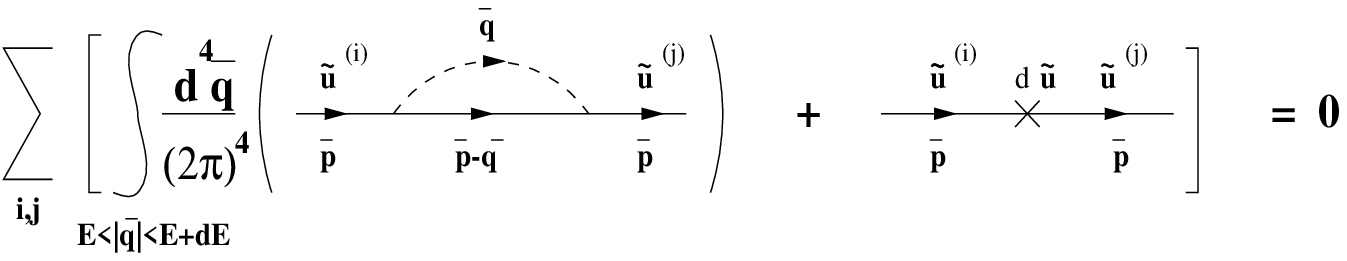} \vskip-0.2cm
\caption{\footnotesize Diagrammatic illustration of the
RGE relation leading to the evolution of quark fields.
This relation means that the two-point Green
functions with truncated external quark legs,
at one-loop level, are the same in the theory with
$E$ cutoff and in the theory with $E\!+\!dE$ cutoff.
Conventions are the same as in previous figures.}
\label{evolq}
\end{figure}

These RGE's can be derived in close analogy with the derivation
of the evolution of scalar fields of the previous Subappendix.
Now, the diagrams (Green functions) of Figs.~\ref{sc2pG} 
and \ref{evolsc} are replaced by 
those of Figs.~\ref{q2pG} and \ref{evolq},
and the scalar field kinetic energy terms in (\ref{kinLep})
are replaced by those of the quark fields.
The Green function of Fig.~\ref{q2pG}, 
with the
incoming ${\tilde u}^{(i)}$ and outgoing ${\tilde u}^{(j)}$
of momentum $p$,
in the framework with UV cutoff $E$, is
\begin{eqnarray}
- {\mathrm{i}} {\Sigma} \left( 
p; E; {\tilde u}^ {(i)}, {\tilde u}^{(j)} \right) &=& 
\frac{{\mathrm{i}}}{64 \pi^2} 
\ln \left( \frac{E^2}{m^2} \right) {p \llap /}
{\Bigg \{} 
 2 \left(1\!+\!{\gamma}_5\right) \sum_{\ell=1}^2 
\left[ {\tilde U}^{(\ell)\dagger} 
{\tilde U}^{(\ell)} \right]_{ji} +
\nonumber\\
&& 
+ \left(1\!-\!{\gamma}_5 \right) \sum_{\ell=1}^2
\left[ {\tilde U}^{(\ell)} {\tilde U}^{(\ell)\dagger}\!+\!
{\tilde D}^{(\ell)} {\tilde D}^{(\ell)\dagger} \right]_{ji} 
{\Bigg \}} \ .
\label{Greenup}
\end{eqnarray}
The Green function with the incoming ${\tilde d}^{(i)}$
and outgoing ${\tilde d}^{(j)}$ of momentum $p$
is obtained from the
above expression by simply exchanging 
${\tilde U}^{(\ell)}\!\leftrightarrow\!{\tilde D}^{(\ell)}$ and
${\tilde U}^{(\ell)\dagger}\!\leftrightarrow\!{\tilde D}^
{(\ell)\dagger}$.
The quark fields evolve according to the ansatz
\begin{equation}
d{\tilde q}^{(k)}(E)_{L,R} 
 =  d f_q(E)_{k\ell}^{(L,R)}
{\tilde q}^{(\ell)}(E)_{L,R} \ , 
\label{ansq}
\end{equation}
where $d {\tilde q}(E)$ generically stands for
${\tilde q}(E\!+\!dE)\!-\!{\tilde q}(E)$
[${\tilde q}^{(k)}\!=\!{\tilde u}^{(k)}, {\tilde d}^{(k)}$],
and subscripts $L$, $R$ denote handedness
of the quark fields: ${\tilde q}_L\!\equiv\!
(1\!-\!{\gamma}_5) {\tilde q} / 2$, 
${\tilde q}_R\!\equiv\!(1\!+\!{\gamma}_5 ) {\tilde q} / 2$.
In complete analogy with the previous Subappendix,
we obtain from this ansatz and from the RGE relation
illustrated in Fig.~\ref{evolq}\footnote{
RGE relation represented by Fig.~\ref{evolq} is analogous
to relation (\ref{kinLep}) represented 
by Fig.~\ref{evolsc}.}
the following relations
for the quark field evolution matrices $d f_u$:
\begin{eqnarray}
\!\!\!\!\!
d f_u(E)_{ij}^{(L)\ast}\!+\!d f_u(E)_{ji}^{(L)} & = &
- \frac{(d \ln E^2)} {32 \pi^2} \sum_{k=1}^2 \left[
{\tilde U}^{(k)} {\tilde U}^{(k)\dagger}\!+\! 
{\tilde D}^{(k)} {\tilde D}^{(k)\dagger} \right]_{ji}(E) ,
\label{fuL}
\\
\!\!\!\!\!
d f_u(E)_{ij}^{(R)\ast}\!+\!d f_u(E)_{ji}^{(R)} & = &
- \frac{2(d \ln E^2)}{32 \pi^2} \sum_{k=1}^2 \left[
{\tilde U}^{(k)\dagger} {\tilde U}^{(k)} \right]_{ji}(E) \ .
\label{fuR}
\end{eqnarray}
The relations for the $d f_d$ evolution matrices of the down-type
sector are obtained from the above by simple exchanges
${\tilde U}^{(\ell)}\!\leftrightarrow\!{\tilde D}^{(\ell)}$ and
${\tilde U}^{(\ell)\dagger}\!\leftrightarrow\!{\tilde D}^
{(\ell)\dagger}$.
Hermitean solution to these relations is
\begin{eqnarray}
\!\!\!\!\!
d f_u(E)_{ij}^{(L)} & = & 
- \frac{(d \ln E^2)}{64 \pi^2} 
\sum_{k=1}^2 \left[
{\tilde U}^{(k)} {\tilde U}^{(k)\dagger}\!+\! 
{\tilde D}^{(k)} {\tilde D}^{(k)\dagger} \right]_{ij}(E) =
d f_d(E)_{ij}^{(L)} ,
\label{fLres}
\\
\!\!\!\!\!
d f_u(E)_{ij}^{(R)} & = & 
- \frac{2(d \ln E^2)}{64 \pi^2} 
\sum_{k=1}^2 \left[
{\tilde U}^{(k)\dagger} {\tilde U}^{(k)} \right]_{ij}(E) ,
\label{fuRres}
\\
\!\!\!\!\!
d f_d(E)_{ij}^{(R)} & = & 
- \frac{2(d \ln E^2) }{64 \pi^2} \sum_{k=1}^2 \left[
{\tilde D}^{(k)\dagger} {\tilde D}^{(k)} \right]_{ij}(E) \ .
\label{fdRres}
\end{eqnarray}
The results (\ref{fLres})-(\ref{fdRres}), in conjuction
with (\ref{ansq}), represent one-loop RGE's for evolution
of the quark fields in the G2HDM, but without the gauge
boson contributions. The latter contributions
are the same as in the MSM
and can be included in (\ref{fLres})-(\ref{fdRres}).

\subsection[]{{\bf One-loop RGE's for the Yukawa coupling 
matrices}}

\begin{figure}[htb]
\mbox{}
\vskip4.5cm\relax\noindent\hskip-0.8cm\relax
\includegraphics{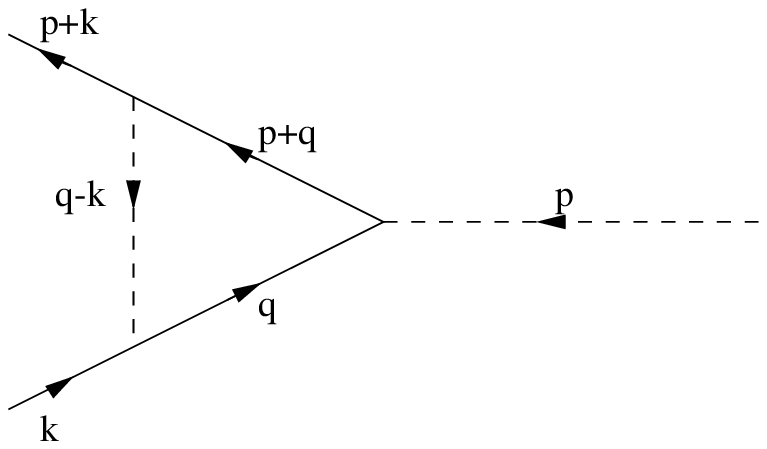} \vskip0.1cm
\caption{\footnotesize One-particle-irreducible (1PI) diagram
contributing to the evolution of the Yukawa coupling
parameters. Conventions are the same is in previous
figures.}
\label{Y3pG}
\end{figure}

To derive these RGE's, we need, in addition to results
of the two previous Subappendices, also
another Green function. It is represented
by the diagram of Fig.~\ref{Y3pG}.
When the external legs there are ${\tilde u}^{(i)}$ 
(incoming, with momentum $k$),
${\tilde u}^{(j)}$ 
(outgoing, with momentum $p\!+\!k$), and ${\phi}_3^{(\ell)}$
[or ${\phi}_4^{(\ell)}$], it turns out that only the
diagram with the {\em charged\/} scalar exchange
contributes, and the resulting truncated three-point Green function,
in the framework with the UV cutoff $E$, is
\begin{eqnarray}
\lefteqn{
G^{(3)}\left( k,p;E; {\tilde u}^{(i)},
{\tilde u}^{(j)}; {\phi}_3^{(\ell)}
\right)  =  - \frac{{\mathrm{i}}}{32 \pi^2 \sqrt{2}}
\ln \left( \frac{E^2}{m^2} \right) \times  }
\nonumber\\
&& 
\,\,\,\,\,\,
\times \sum_{r=1}^{2}
{\Bigg \{} (1\!+\!{\gamma}_5) \left[ 
{\tilde D}^{(r)} {\tilde D}^{(\ell)\dagger} {\tilde U}^{(r)} 
\right]_{ji} +
(1\!-\!{\gamma}_5) \left[
{\tilde U}^{(r)\dagger} {\tilde D}^{(\ell)} {\tilde D}^{(r)\dagger} 
\right]_{ji} {\Bigg \}} \ .
\label{3ptupGreen}
\end{eqnarray}
The corresponding Green function with the down-type quark
external legs is obtained from the above by the exchanges
${\tilde U}^{(s)}\!\leftrightarrow\!{\tilde D}^{(s)}$ and
${\tilde U}^{(s)\dagger}\!\leftrightarrow\!{\tilde D}^
{(s)\dagger}$.

\begin{figure}[htb]
\mbox{}
\vskip2.5cm\relax\noindent\hskip1.1cm\relax
\includegraphics{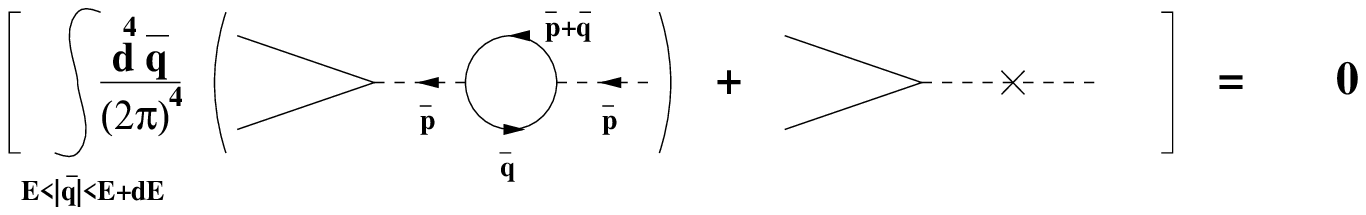} \vskip-0.1cm
\caption{\footnotesize Cancellation of contributions
from the quark loop
(one-particle-reducible -- 1PR) with those of the 
scalar field renormalizations in the kinetic
energy term of the scalars, for the energy cutoff
interval $(E,E\!+\!dE)$.}
\label{canc1}
\end{figure}

\begin{figure}[htb]
\mbox{}
\vskip2.5cm\relax\noindent\hskip1.1cm\relax
\includegraphics{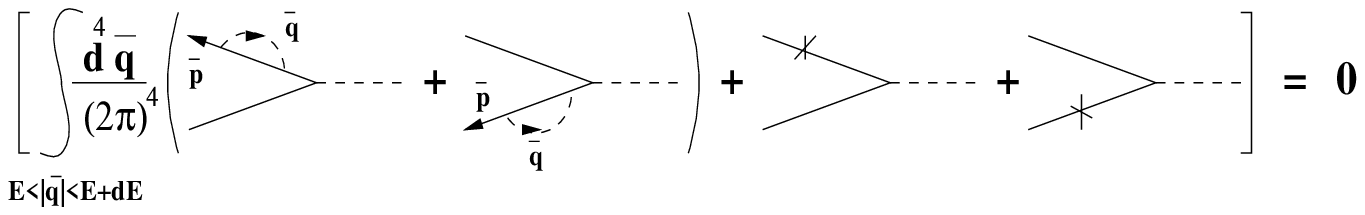} \vskip-0.1cm
\caption{\footnotesize Cancellation of contributions
from the scalar exchange on the quark legs
(1PR) with those of the 
quark field renormalizations in the kinetic
energy term of the quarks, for the energy cutoff
interval $(E,E\!+\!dE)$.}
\label{canc2}
\end{figure}

\begin{figure}[htb]
\mbox{}
\vskip5.3cm\relax\noindent\hskip1.2cm\relax
\includegraphics{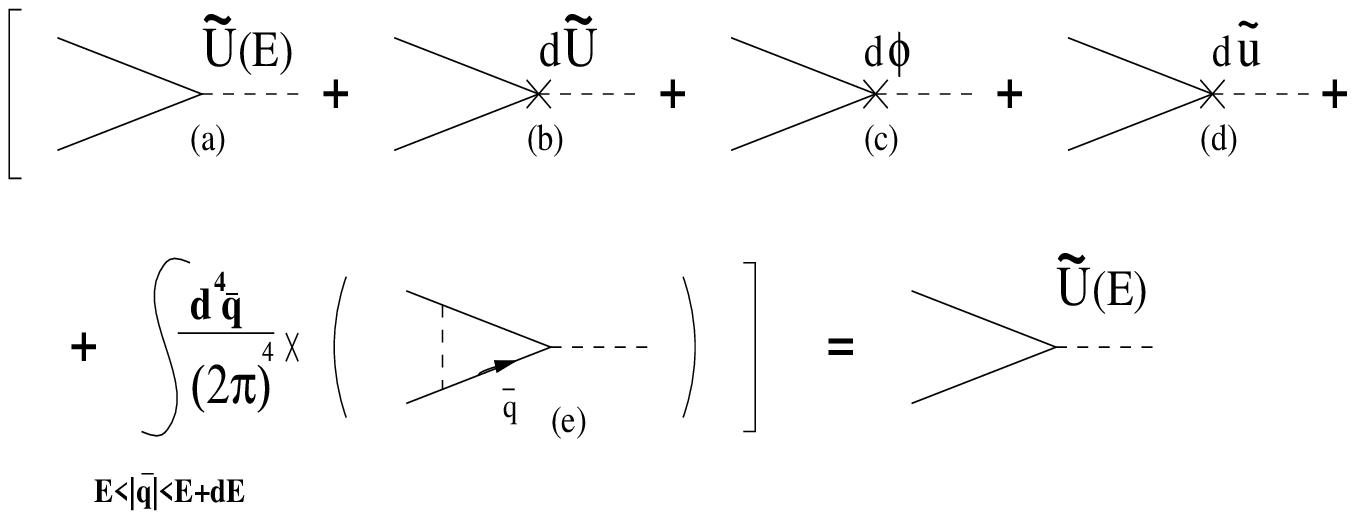} \vskip-0.1cm
\caption{\footnotesize Diagrammatic representation of the
RGE for the up-type Yukawa matrix ${\tilde U}$. Only
the 1PI scalar exchange [(e)]
and the effects of the renormalizations
of the Yukawa matrix, of the scalar fields and the quark fields
in the Yukawa couplings [(b), (c), (d), respectively] 
contribute when the cutoff is changed
from $E$ (RHS) to $E\!+\!dE$ (LHS). 
Note that $d{\tilde U}$ 
stands for ${\tilde U}(E\!+\!dE)\!-\!{\tilde U}(E)$,
etc. The contributions of the gauge boson exchanges were
not considered in the Figure.}
\label{Upict}
\end{figure}

Now, the one-loop RGE's for the Yukawa matrices are
obtained in analogy with the reasoning leading, in
the case of two-point scalar Green functions, to the
RGE relation (\ref{kinLep}) [cf.~Fig.~\ref{evolsc}].
It is straightforward to check that the contribution
of the quark loops in the scalar external leg cancel
the contributions coming from the renormalizations
of the scalar fields in the kinetic energy terms
of the scalars -- this is illustrated in Fig.~\ref{canc1}.
Furthermore, it can be checked
that the contributions of the scalar
exchanges on the external quark legs cancel the
contributions coming from the renormalizations
of the quark fields in the kinetic energy terms of
the quarks -- illustrated in Fig.~\ref{canc2}.
All in all, the 1PR 
one-loop contributions are canceled by the
contributions of field renormalizations in the kinetic energy
terms. Therefore, the only one-loop terms contributing
to evolution of the ${\tilde U}^{(k)}$ Yukawa
matrices are those depicted in Fig.~\ref{Upict}. 
The three
diagrams with crosses there correspond to contributions
of the following changes {\em in the Yukawa coupling} terms:
\begin{itemize}
\item
Yukawa matrix change (renormalization)
$d {\tilde U}^{(k)}$ [$\equiv {\tilde U}^{(k)}(E\!+\!dE)\!-\!
{\tilde U}^{(k)}(E)$] -- Fig.~\ref{Upict}(b);
\item 
the scalar field renormalization $d {\tilde {\phi}}^{(k)}_s$
[$\equiv {\tilde {\phi}}^{(k)}_s(E\!+\!dE)\!-\!
{\tilde {\phi}}^{(k)}_s(E)$] -- Fig.~\ref{Upict}(c);
\item
the quark field renormalization $d {\tilde u}^{(i)}$
and $d {\tilde u}^{(j)}$ -- Fig.~\ref{Upict}(d).
\end{itemize}
Figure \ref{Upict} is a diagrammatical 
representation of the physical
requirement that the three-point (quark-antiquark-scalar)
Green function, at one-loop level, in the theory
with the cutoff $E+dE$ [left side of Fig.~\ref{Upict}: 
$({\mathrm{a}})\!+\!\ldots\!+\!({\mathrm{e}})$] 
be the same as in the theory 
with the slightly lower cutoff $E$
(right side).

Using the results of this and the previous Subappendices,
we can then write down the one-loop RGE for
${\tilde U}^{(k)}$ corresponding to Fig.~\ref{Upict},
at the right-handed component [$ \propto\!(1\!+\!{\gamma}_5)$]
of the three-point Green function
\begin{eqnarray}
\lefteqn{
{\tilde U}^{(k)}_{ji} + d{\tilde U}^{(k)}_{ji}
+ \frac{1}{32 \pi^2} \left( d \ln E^2 \right)
{\Bigg \{} 
- N_{\mathrm{c}} \sum_{\ell=1}^2 {\mathrm{Tr}}
\left[ {\tilde U}^{(k)} {\tilde U}^{(\ell)\dagger} +
 {\tilde D}^{(\ell)} {\tilde D}^{(k)\dagger} \right]
{\tilde U}^{(\ell)} }
\nonumber\\
&&
 - \frac{1}{2} \sum_{\ell=1}^2 \left[
\left( {\tilde U}^{(\ell)} {\tilde U}^{(\ell)\dagger} 
+
{\tilde D}^{(\ell)} {\tilde D}^{(\ell)\dagger} \right)
{\tilde U}^{(k)} + 
2 {\tilde U}^{(k)} {\tilde U}^{(\ell) \dagger}
{\tilde U}^{(\ell)} \right]
\nonumber\\
&&+ 2 \sum_{\ell=1}^2  
\left[ {\tilde D}^{(\ell)} 
{\tilde D}^{(k)\dagger} {\tilde U}^{(\ell)} 
\right]
{\Bigg \}}_{ji}
 =  {\tilde U}^{(k)}_{ji} \ .
\label{RGEU1}
\end{eqnarray}
Taking the left-handed component of the three-point Green function
results in the Hermitean conjugate of Eq.~(\ref{RGEU1}), i.e., in
an equivalent relation.
The first sum on the left ($\propto\!N_{\rm{c}}$)
of Eq.~(\ref{RGEU1}) corresponds to Fig.~\ref{Upict}(c) 
[cf.~Eqs.~(\ref{RGEphi1Y})-(\ref{RGEPhiY})], 
the second sum to Fig.~\ref{Upict}(d) [cf.~Eqs.~(\ref{fLres}), 
(\ref{fuRres})], 
and the third sum to Fig.~\ref{Upict}(e) 
[cf.~Eq.~(\ref{3ptupGreen})].
The left-handed part of the Green function yields just the
Hermitean conjugate of the above matrix relation. The
analogous consideration of the three-point Green functions
with the down-type external quark legs ${\tilde d}^{(i)}$
and ${\tilde d}^{(j)}$ gives relations which can be
obtained from the above relation again by the exchanges
${\tilde U}^{(s)}\!\leftrightarrow\!{\tilde D}^{(s)}$ and
${\tilde U}^{(s)\dagger}\!\leftrightarrow\!{\tilde D}
^{(s)\dagger}$.
These relations can be rewritten in a more conventional
form
\begin{eqnarray}
16 \pi^2 \frac{d}{d \ln E} {\tilde U}^{(k)}(E) & = &
{\Bigg \{} 
N_{\mathrm{c}} \sum_{\ell=1}^2
{\mathrm{Tr}} \left[ {\tilde U}^{(k)} 
{\tilde U}^{(\ell)\dagger} +
 {\tilde D}^{(\ell)} {\tilde D}^{(k)\dagger} \right]
{\tilde U}^{(\ell)}
\nonumber\\
&&+ \frac{1}{2} \sum_{\ell=1}^2 \left[
{\tilde U}^{(\ell)} {\tilde U}^{(\ell)\dagger} +
{\tilde D}^{(\ell)} {\tilde D}^{(\ell)\dagger} 
\right] {\tilde U}^{(k)}
\nonumber\\
&&+ {\tilde U}^{(k)} \sum_{\ell=1}^2
{\tilde U}^{(\ell)\dagger} {\tilde U}^{(\ell)} 
-2 \sum_{\ell=1}^2 \left[
{\tilde D}^{(\ell)} {\tilde D}^{(k)\dagger}
{\tilde U}^{(\ell)} \right] {\Bigg \}} \ ,
\label{RGEU2}
\end{eqnarray}
and an analogous RGE for ${\tilde D}^{(k)}$.
These RGE's still don't contain one-loop effects of
exchanges of gauge bosons. However, since the couplings
of quarks and the Higgs doublets to the gauge bosons are
identical to those in the usual MSM, 2HDM(I) and 2HDM(II), 
their contributions
on the right of the above RGE's are identical to those in 
these theories. Therefore, the final form of the one-loop
RGE's for the Yukawa matrices in the general 2HDM now reads
\begin{eqnarray}
\lefteqn{
16 \pi^2 \frac{d}{d \ln E} {\tilde U}^{(k)}(E)  = 
{\Bigg \{} 
N_{\mathrm{c}} \sum_{\ell=1}^2
{\mathrm{Tr}} \left[ {\tilde U}^{(k)} 
{\tilde U}^{(\ell)\dagger} +
 {\tilde D}^{(\ell)} {\tilde D}^{(k)\dagger} \right]
{\tilde U}^{(\ell)}}
\nonumber\\
&&+\frac{1}{2} \sum_{\ell=1}^2 \left[
{\tilde U}^{(\ell)} {\tilde U}^{(\ell)\dagger} +
{\tilde D}^{(\ell)} {\tilde D}^{(\ell)\dagger} 
\right] {\tilde U}^{(k)}
+ {\tilde U}^{(k)} \sum_{\ell=1}^2
{\tilde U}^{(\ell)\dagger} {\tilde U}^{(\ell)}
\nonumber\\ 
&&-2 \sum_{\ell=1}^2 \left[
{\tilde D}^{(\ell)} {\tilde D}^{(k)\dagger}{\tilde U}^{(\ell)}
\right]  - A_U {\tilde U}^{(k)}
{\Bigg \}} \ ,
\label{RGEUk}
\end{eqnarray}
\begin{eqnarray}
\lefteqn{
16 \pi^2 \frac{d}{d \ln E} {\tilde D}^{(k)}(E)  = 
{\Bigg \{} 
N_{\mathrm{c}} \sum_{\ell=1}^2
{\mathrm{Tr}} \left[ {\tilde D}^{(k)} 
{\tilde D}^{(\ell)\dagger} +
 {\tilde U}^{(\ell)} {\tilde U}^{(k)\dagger} \right]
{\tilde D}^{(\ell)}}
\nonumber\\
&&+\frac{1}{2} \sum_{\ell=1}^2 \left[
{\tilde U}^{(\ell)} {\tilde U}^{(\ell)\dagger} +
{\tilde D}^{(\ell)} {\tilde D}^{(\ell)\dagger} 
\right] {\tilde D}^{(k)}
+ {\tilde D}^{(k)} \sum_{\ell=1}^2
{\tilde D}^{(\ell)\dagger} {\tilde D}^{(\ell)}
\nonumber\\ 
&&-2 \sum_{\ell=1}^2 \left[
{\tilde U}^{(\ell)} {\tilde U}^{(k)\dagger}{\tilde D}^{(\ell)}
\right]  - A_D {\tilde D}^{(k)}
{\Bigg \}} \ ,
\label{RGEDk}
\end{eqnarray}
where the functions $A_U$ and $A_D$, characterizing the
contributions of the gauge boson exchanges, are 
gauge independent and are the
same as in the MSM, 2HDM(I) and 2HDM(II)
\begin{equation}
A_U = 3 \frac{(N_{\mathrm{c}}^2 - 1)}
{ N_{\mathrm{c}} } g_3^2 + \frac{9}{4} g_2^2 +
\frac{17}{12} g_1^2 \ ,
\qquad
A_D  =  A_U - g_1^2 \ ,
\label{AUAD}
\end{equation}
and the gauge coupling parameters $g_j$ satisfy the one-loop
RGE's
\begin{equation}
16 \pi^2 \frac{d}{d \ln E} g_j= - C_j g_j^3 \ ,
\label{RGEgj}
\end{equation}
with the coefficients $C_j$ being those for the 2HDM's ($N_H\!=\!2$)
\begin{equation}
C_3 = \frac{1}{3}(11 N_{\mathrm{c}} - 2 n_q) \ ,
\quad C_2 = 7 - \frac{2}{3} n_q \ , \quad
C_1 = - \frac{1}{3} - \frac{10}{9} n_q \ .
\label{Cjs}
\end{equation}
Here, $n_q$ is the number of effective quark flavors --
\eg{}, for $E\!>\!m_t$ we have $n_q\!\approx\!6$; for 
$m_b\!<\!E\!<\!m_t$
we have $n_q\!\approx\!5$, etc.

This completes the derivation of the one-loop RGE's.

\end{appendix}

\end{document}